\documentclass{aa}

\usepackage{graphicx,natbib}
\usepackage{floatrow}
\usepackage{subfigure}
\usepackage{amsmath,amssymb}
\usepackage[T1]{fontenc}

\usepackage{txfonts}

\usepackage{hyperref}

\hypersetup{
   colorlinks=true,
   citecolor=blue,
   linkcolor=blue,
   urlcolor=black
   }

\def\brunt{Brunt-V\"ais\"al\"a }

\def\gl581{Gl\,581\,c}
\def\hd85{HD\,85512\,b}

\def\ssb{\sigma}
\def\kB{k_\mathrm{\,B}}

\def\msun{ M_\odot}
\def\lsun{ L_\odot}

\def\rearth{ R_\oplus}
\def\mearth{ M_\oplus}

\def\co2{CO$_2$}
\def\h2o{H$_2$O}
\def\ch4{CH$_4$}
\def\N2{N$_2$}
\def\qvap{q_{\mathrm{v}}}
\def\ql{q_{\mathrm{l}}}
\def\qsat{q_{\mathrm{s}}}
\def\mvap{m^{\star}_{\mathrm{v}}}
\def\dmvap{\dot{m}^{\star}_{\mathrm{v}}}
\def\tauvap{\tau_{\mathrm{v}}^\nu}
\def\kapvap{\kappa_{\mathrm{v}}^\nu}
\def\nmix{N_{\mathrm{c}}}
\def\tauc{\tau_\mathrm{c}}

\def\Rp{R_\mathrm{\,p}}
\def\Mp{M_\mathrm{\,p}}
\def\kp{k_{\,2,\mathrm{p}}}

\def\dtp{\Delta t_\mathrm{\,p}}
\def\op{\Omega}
\def\ep{\varepsilon_\mathrm{\,p}}

\def\alb{\bar{A}}
\def\lalb{A}
\def\redist{\eta}

\def\Ms{M_\star}

\def\Ls{L_\star}
\def\Fs{F_\star}
\def\mus{\mu_\star}

\def\Fgeo{\phi_\mathrm{g}}

\def\grav{g}
\def\cp{c_\mathrm{p}}

\def\CLF{f_\mathrm{c}}

\def\rhoi{\rho_\mathrm{ice}}
\def\Lro{L_\mathrm{Ro}}
\def\Ro{\mathcal{L}}
\def\coriolis{f}
\def\Nbv{N}
\def\kapth{K_\mathrm{th}}

\def\hp{H}
\def\tequ{T_\mathrm{eq}}
\def\ltequ{T_\mathrm{eq,\mu_\star}}
\def\tsurf{\bar{T}_\mathrm{s}}
\def\ltsurf{T_\mathrm{s}}
\def\ps{p_\mathrm{s}}
\def\pmelt{p_\mathrm{melt}}
\def\ptriple{p_0}
\def\ttriple{T_0}

\def\tadv{\tau_\mathrm{adv}}
\def\trad{\tau_\mathrm{rad}}

\def\d{\mathrm{d}}

\newcommand{\balign}[1]{
\begin{align}
#1
\end{align}}

\newcommand{\fracl}[2]{\raisebox{0.4ex}{$#1$} / \raisebox{-0.7ex}{$#2$}}

\newcommand{\eq}[1]{Eq.\,(\ref{#1})}
\newcommand{\eqs}[2]{Eqs.\,(\ref{#1}) and (\ref{#2})}
\newcommand{\fig}[1]{Fig.\,\ref{#1}}
\newcommand{\figs}[2]{Figs\,\ref{#1} and \ref{#2}}
\newcommand{\sect}[1]{Sect.\,\ref{#1}}

\newcommand{\tab}[1]{Table\,\ref{#1}}

\newcommand{\dd}[2]{\frac{\mathrm{d} \!\! \ #1}{\mathrm{d}\!\! \ #2}}

\titlerunning{3D climate modeling of close-in land planets}
\authorrunning{Leconte et al.}

\begin{document}

\title{
3D climate modeling of close-in land planets:\\
Circulation patterns, climate moist bistability and habitability
}

\author{J\'er\'emy Leconte\inst{1} \and Francois Forget\inst{1} \and Benjamin Charnay\inst{1} \and  Robin Wordsworth\inst{2}  \and Franck Selsis \inst{3,4}  \and Ehouarn Millour\inst{1}}

\institute{LMD, Institut Pierre-Simon Laplace, Universit\'e P. et M. Curie, BP99, 75005, Paris, France\\
\email{jeremy.leconte@lmd.jussieu.fr}
\and
Department of Geological Sciences, University of Chicago, 5734 S Ellis Avenue, Chicago, IL 60622, USA
\and
Universit\'e de Bordeaux, Observatoire Aquitain des Sciences de l'Univers, BP 89, 33271 Floirac Cedex, France
\and
CNRS, UMR 5804, Laboratoire d'Astrophysique de Bordeaux, BP 89, 33271 Floirac Cedex, France
}

\date{Accepted 27 March 2013}

\offprints{J. Leconte}

\abstract{
The inner edge of the classical habitable zone is often defined by the critical flux needed to trigger the runaway greenhouse instability. This 1D notion of a critical flux, however, may not be so relevant for inhomogeneously irradiated planets, or when the water content is limited (land planets).\\
Here, based on results from our 3D global climate model, we present general features of the climate and large scale circulation on close-in terrestrial planets. We find that the circulation pattern can shift from super-rotation to stellar/anti stellar circulation when the equatorial Rossby deformation radius significantly exceeds the planetary radius, changing the redistribution properties of the atmosphere.
Using analytical and numerical arguments, we also demonstrate the presence of systematic biases between mean surface temperatures or temperature profiles predicted from either 1D or 3D simulations.\\
Including a complete modeling of the water cycle, we further demonstrate that for land planets closer than the inner edge of the classical habitable zone, two stable climate regimes can exist. One is the classical runaway state where all the water is vaporized, and the other is a collapsed state where water is captured in permanent cold traps. We identify this "moist" bistability as the result of a competition between the greenhouse effect of water vapor and its condensation on the night side or near the poles, highlighting the dynamical nature of the runaway greenhouse effect. We also present synthetic spectra showing the observable signature of these two states.\\
Taking the example of two prototype planets in this regime, namely Gl\,581\,c and HD\,85512\,b, we argue that depending on the rate of water delivery and atmospheric escape during the life of these planets, they could accumulate a significant amount of water ice at their surface. If such a thick ice cap is present, various physical mechanisms observed on Earth (e.g. gravity driven ice flows, geothermal flux) should come into play to produce long-lived liquid water at the edge and/or bottom of the ice cap.
Consequently, the habitability of planets at smaller orbital distance than the inner edge of the classical habitable zone cannot be ruled out. Transiting planets in this regime represent promising targets for upcoming exoplanet characterization observatories like EChO and JWST.
}


\maketitle

\section{Beyond the runaway greenhouse limit}
\label{sec:intro}

Because of the bias of planet characterization observatories toward strongly irradiated objects on short orbits, planets identified to be near the inner edge of the habitable zone represent valuable targets. This however raises numerous questions concerning our understanding of the processes delimiting this inner edge.

For a planet like the Earth, which from the point of view of climate can be considered as an aqua planet\footnote{i.e. a planet whose surface is largely covered by connected oceans that have a planet-wide impact on the climate. Not to be confused with an ocean planet for which, in addition, water must represent a significant fraction of the bulk mass \citep{LSS04}.}, a key constraint is provided by the positive feedback of water vapor on climate. When the insolation of the planet is increased, more water vapor is released in the atmosphere by the hotter surface, increasing the greenhouse effect. If on Earth today this effect is balanced by other processes, in particular the increase of infrared emission by the surface with temperature, it has been shown that above a critical absorbed stellar flux, the planet cannot reach radiative equilibrium balance and is heated until the surface can radiate at optical wavelength around a temperature of 1400K \citep{KPA84,Kas88}.
 All the water at the surface is vaporized.  The planet is in a runaway greenhouse state. Roughly speaking, this radiation limit stems from the fact that the surface must increase its temperature to radiate more, but the amount of water vapor in the troposphere, and thus the opacity of the latter, increases exponentially with this surface temperature (because of the Clausius-Clapeyron relation). Then the thermal emission of the opaque atmosphere is emitted at the level at which the optical depth is close to unity, a level whose temperature reaches an asymptotic value, hence the limited flux\footnote{A similar but slightly subtler limit stems from considering the radiative/thermodynamic equilibrium of the stratosphere \citep{Kom67,Ing69,NHA92}, but it is less constraining in the case considered.} \citep{NHA92,Pie10}.

While large uncertainties remain regarding the radiative effect of water vapor continuum and the nonlinear, three-dimensional effect of dynamics and clouds, the critical value of the average absorbed stellar radiation, $(1-\alb)\Fs/4$ (where $\Fs$ is the stellar flux at the substellar point and $\alb$ is the planetary bond albedo), should conservatively range between 240\,W/m$^2$ (the mean absorbed stellar radiation on Earth) and 350\,W/m$^2$ \citep{AAS11}. Terrestrial planets emitting such (low) IR fluxes will be difficult to characterize in the near future.

Fortunately, the runaway greenhouse limit discussed above is not as fundamental as it may appear. Indeed, the theory outlined above assumes that a \textit{sufficient}\footnote{Meaning that, if fully vaporized, the water reservoir must produce a surface pressure which is higher than the pressure of the critical point of water (220bars).} reservoir of liquid water is available \textit{everywhere} on the surface \citep{SKL07}. While it would seem from 1D modeling that liquid water would disappear at a lower surface temperature for a planet with a limited water inventory, it goes the \textit{opposite} way in 3D. Global climate models show that such "land planets" can have a wider habitable zone \citep{AAS11}. Near the inner edge, this property stems from the fact that atmospheric transport of water from warm to cold regions of the planets is not counteracted by large-scale surface runoff \citep{ANK05}. Heavily irradiated regions are then drier and can emit more IR flux than the limit found for a saturated atmosphere.

In principle, it is thus possible to find planets absorbing and re-emitting more flux than the runaway greenhouse threshold and for which liquid/solid water can be thermodynamically stable at the surface, provided that they accreted limited water supplies (or lost most of them). As will be shown hereafter, this could be the case for \gl581 \citep{UBD07,MBF09,FBD11} and \hd85 \citep{PLS11}, the only two confirmed planets having masses compatible with a terrestrial planet ($\Mp\lesssim7-8 \mearth$; although defining such a mass limit is questionable), lying beyond the inner edge of the classical habitable zone but receiving a moderate stellar flux, $\Fs/4<1000$\,W/m$^2$.
Note that these planets lie even beyond the inner edge of the limits set by \citet{AAS11} for land planets.

\begin{table*}[htbp]
\centering
\caption{Standard parameters used in the climate simulations.}
\begin{tabular}{llll}
\hline \hline
Planet name & & \gl581& \hd85\\
Stellar luminosity & $\Ls$ [$\lsun$] & 0.0135 & 0.126\\
Stellar mass & $\Ms$ [$\msun$] & 0.31 & 0.69\\
Orbital semi-major axis & $a$ [AU] & 0.073 &0.26\\
Orbital rotation rate & $\Omega_\mathrm{orb}$ [s$^{-1}$] & 5.59$\times10^{-6}$ & 1.25$\times10^{-6}$ \\
Orbital eccentricity & $e$ & 0-0.05 & 0-0.11\\
Obliquity & $\ep$ & 0 & --\\
Tidal resonance & $n$ &  1, 3:2& -- \\
atmospheric pressure & $p_{s}$ [bar] &  0.1-10 & --\\
Mass$^{\dag}$ & $\Mp$ [$\mearth$] & 6.25 & 4.15 \\
Radius$^\ddag$ & $\Rp$ [$\rearth$] & 1.85 & 1.60\\
Surface gravity & $\grav$ [m s$^{-2}$] & 18.4 & 15.8\\
Surface albedo & $A_\mathrm{s}$ & 0.3 &--\\
Ice albedo & $A_\mathrm{ice}$ & 0.3-0.55 &--\\
Surface roughness & $z_0$ [m] & $1\times10^{-2}$ &--\\
Specific heat capacity & $\cp$ [J\,K$^{-1}$kg$^{-1}$] & 1003.16 & --\\
Condensation nuclei & $\nmix$ [kg$^{-1}$] & $1\times10^5$ &--\\
\hline \hline\\
\multicolumn{4}{l}{$^\dag$ This is the mass used in simulations. It is inferred from the $\Mp\sin i$} \\
\multicolumn{4}{l}{measured by radial velocity by assuming $i=60^\circ$. This inclination}\\
\multicolumn{4}{l}{seems favored by recent observations of a disk around Gl\,581.}\\
\multicolumn{4}{l}{$^\ddag$ Inferred from the mass assuming Earth density.}
\end{tabular}\label{tab:params}
\end{table*}


However, for such close-in planets whose rotation states have been strongly affected by tidal dissipation and which are either in a state of synchronous, pseudo-synchronous\footnote{Note that the stability of the pseudo-synchronous rotation state is still debated for terrestrial planets \citep{ME12}} or resonant rotation with a near zero obliquity \citep{GP66,HLB11,MBE12}, efficient and permanent cold traps present on the dark hemisphere or at the poles could irreversibly capture all the available water in a permanent ice cap.
To tackle this issue, we have performed several sets of global climate model simulations for prototype land planets with the characteristic masses and (strong) incoming stellar fluxes of \gl581 and \hd85. 3D climate models are necessary to assess the habitability of land planets in such scenarios because horizontal inhomogeneities in impinging flux and water distribution are the key features governing the climate. This may explain why this scenario has received little attention so far.

One of the difficulties lies in the fact that, as will be extensively discussed here, there is no precise flux limit above which a runaway greenhouse state is inevitable. Instead, for a wide range of parameters, two stable equilibrium climate regimes exist because of the very strong positive feedback of water vapor. The actual regime is determined not only by the incoming stellar flux and spectrum, or by the planet's atmosphere, but also by the amount of water available and its initial distribution.
Consequently, instead of trying to cover the whole parameter space, here we have performed simulations for two already detected planets. We will show that qualitatively robust features appear in our simulations, and argue that the processes governing the climate discussed here are quite general.

In \sect{sec:model} we introduce the various components of our numerical climate model and we present the basic climatology of our model in a completely dry case (\sect{sec:1d}). We show that the atmospheric circulation can settle into two different regimes - namely super-rotation or stellar/anti-stellar circulation - and that the transition of one regime to the other occurs when the equatorial Rossby deformation radius significantly exceeds the planetary radius. We also show that strong discrepancies can occur between 1D and globally averaged 3D results. Then we show that when water is present, the climate regime reached depends on the initial conditions and can exhibit a bistability (\sect{sec:coldtrap}). We argue that this bistability is a dynamical manifestation of the interplay between the runaway greenhouse instability on the hot hemisphere and the cold trapping arising either on the dark side or near the poles. In \sect{sec:ice_limits}, we discuss the long-term stability of the different regimes found here and show that, if atmospheric escape is efficient, it could provide a stabilizing feedback for the climate. In \sect{sec:icecap}, we suggest that if the planet has been able to accumulate a large amount of ice in its cold traps such that ice draining by glaciers is significant, liquid water could be present at the edge of the ice cap where glaciers melt. We also discuss the possible presence of subsurface water under a thick ice cap. Finally, in \sect{sec:obs} we present synthetic spectra and discuss the observable signatures of the two climate regimes discussed above.

\section{Method}
\label{sec:model}

Our simulations have been performed with an upgraded version of the LMD generic Global Climate Model (GCM) specifically developed for the study of extrasolar planets \citep{WFS10,WFS11,SWF11} and paleoclimates \citep{WFM13,FWM13}. The model uses the 3D dynamical core of the LMDZ 3 GCM \citep{HMB06}, based on a finite-difference formulation of the primitive equations of geophysical fluid dynamics. A spatial resolution of $64\times48\times20$ in longitude, latitude and altitude is used for the simulations.

\subsection{radiative transfer}

The method used to produce our radiative transfer model is similar to \citet{WFS11}. For a given mixture of atmospheric gases, here \N2 with 376\,ppmv of \co2 and a variable amount of water vapor, we computed high resolution spectra over a range of temperatures and pressures using the HITRAN 2008 database \citep{RGB09}. Because the \co2 concentration depends on the efficiency of complex weathering processes, it is mostly unconstrained for hot synchronous land planets. The choice of an Earth-like concentration is thus arbitrary but our prototype atmosphere should be representative of a background atmosphere with a low greenhouse effect. For this study we used temperature, pressure grids with values $T = \{110,170,...,710\}\, \mathrm{K}, p = \{10^{-3},10^{-2},...,10^{5}\}\, \mathrm{mbar}$. The \h2o volume mixing ratio could vary in the range $\{10^{-7},10^{-6},...,1\}$. \h2o lines were truncated at 25\,cm$^{-1}$, while the water vapor continuum was included using the CKD model \citep{CKD89}. We also account for opacity due to \N2-\N2 collision induced absorption \citep{BF86,RGR11}.

The correlated-k method was then used to produce a smaller database of coefficients suitable for fast calculation in a GCM. Thanks to the linearity of the Schwarzschild equation of radiative transfer, the contribution of the thermal emission and downwelling stellar radiation can be treated separately, even in the same spectral channel. Hence, we do not assume any spectral separation between the stellar and planetary emission wavelengths. For thermal emission, the model uses 19 spectral bands, and the two-stream equations are solved using the hemispheric mean approximation \citep{TMA89}. Infrared scattering by clouds is also taken into account. Absorption and scattering of the downwelling stellar radiation is treated with the $\delta$-Eddington approximation within 18 bands. Sixteen points are used for the $\bar{g}$-space integration, where $\bar{g}$ is the cumulated distribution function of the absorption data for each band.

As in \citet{WFS10}, the emission spectrum of our M dwarf, Gl\,581, was modeled using the Virtual Planet Laboratory AD Leo data \citep{SKM05}. For the K star HD\,85512, we used the synthetic spectrum of a K dwarf with an effective temperature of 4700\,K, $\grav=10^{2.5}$ m\,s$^{-2}$ and $\left[\mathrm{M/H}\right]=-0.5$\,dex from the database of \citet{AHS00}.

\subsection{water cycle}

In the atmosphere, we follow the evolution of water in its vapor and condensed phase. These tracers are advected by the dynamical core, mixed by turbulence and dry and moist convection.

Much care has been devoted to develop a robust and numerically stable water cycle scheme that is accurate both in the trace gas (water vapor mass mixing ratio $\qvap\ll 1$) and dominant gas ($\qvap\sim 1$) limit.
In particular, the atmospheric mass and surface pressure variation (and the related vertical transport of tracers through pressure levels) due to any evaporation/sublimation/condensation of water vapor is taken into account with a scheme similar to the one developed by \citet{FML06} for \co2 atmospheres.

Cloud formation is treated using the prognostic equations of \citet{HL91}. For each column and level, this scheme provides the cloud fraction, $\CLF$, and the mass mixing ratio amount of condensed water, $\ql$, as a function of $\qvap$ and the saturation vapor mass mixing ratio, $\qsat$. In addition, when part of a column reaches both 100\% saturation and a super-moist-adiabatic lapse rate, moist convective adjustment is performed following \citet{MW67} and the cloud fraction is set to unity. In the cloudy part of the sky, the condensed water is assumed to form a number $\nmix$ of cloud particles per unit mass of moist air. This quantity, which represents the number density per unit mass of activated cloud condensation nuclei, is assumed to be uniform throughout the atmosphere \citep{FWM13}. Then, the effective radius of the cloud particles is given by
$r_\mathrm{eff}=\left(\fracl{3\,\ql}{4\,\pi\,\rho_\mathrm{c}\,\nmix}\right)^{1/3}$, where $\rho_\mathrm{c}$ is the density of condensed water (10$^3$\,kg/m$^3$ for liquid and 920\,kg/m$^3$ for ice). Precipitation is computed with the scheme of \citet{BLB95}. As this scheme explicitly takes into account the dependence on gravity, cloud particle radii and background air density, it should remain valid for a wide range of situations. Finally, the total cloud fraction of the column is assumed to be the cloud fraction of the level with the thickest cloud, and radiative transfer is computed both in the cloudy and clear sky regions. Fluxes are then linearly weighted between the two regions.

On the surface, ice can also have a radiative effect by linearly increasing the albedo of the ground to $A_\mathrm{ice}$ (see \tab{tab:params}; in the present version of the code, both surface and ice albedos are independent of the wavelength) until the ice surface density exceeds a certain threshold (here 30\,kg\,m$^{-2}$). In the general case, the values taken for both of these parameters can have a significant impact on the mean radiative balance of a planet. In addition, providing an accurate ice mean albedo can prove difficult, as it depends on the ice cover, light incidence, state of the ice/snow, but also on the stellar spectrum (ice absorbing more efficiently infrared light; \citealt{JH12}). However, in our case of a strongly irradiated, spin synchronized land planet where water is in limited supply, most of the day side is above 0$^\circ$C, and ice can form only in regions receiving little or no visible flux. As a result, and as confirmed by our tests, the value chosen for $A_\mathrm{ice}$ has a negligible impact on the results presented hereafter.



\section{Climate on synchronous planets: what 1D cannot tell you}
\label{sec:1d}

So far, most studies concerned with the determination of the limits of the habitable zone have used 1D calculations. While these single column models can provide reasonable answers for planets with (very) dense atmospheres and/or a rapid rotation that limits large-scale temperature contrasts, we will show here that only a three-dimensional model can treat the case of synchronized or slowly rotating exoplanets properly. To that purpose we will quantify both analytically and numerically the discrepancies that can arise between 1D and 3D models.

\subsection{Equilibrium temperatures vs physical temperatures}\label{sec:analytical}

When considering the energetic balance of an object, one of the first quantities coming into mind is the equilibrium temperature, i.e. the temperature that a blackbody would need to have to radiate the absorbed flux. If one assumes local thermal equilibrium of each atmospheric column, one can define the \textit{local} equilibrium temperature
\balign{\label{loceqtemp}
\ltequ\equiv\left(\frac{(1-\lalb)\,\Fs\,\mus}{\ssb}\right)^{1/4},}
where $\Fs$ is the stellar flux at the substellar point at the time considered, $\lalb$ is the local bond albedo, $\mus$ is the cosine of the stellar zenith angle (assumed equal to zero when the star is below the horizon), and $\ssb$ is the Stefan-Boltzmann constant.

Relaxing the local equilibrium condition but still assuming global thermal equilibrium of the planet, one can define the \textit{effective} equilibrium temperature
\balign{\label{eqtemp}
\tequ\equiv \left(\frac{1}{S\, \ssb} \int(1-\lalb)\,\Fs\,\mus \d S\right)^{1/4}\equiv\left(\frac{(1-\alb)\,\Fs}{4\, \ssb}\right)^{1/4},}
where $S$ is the planet's surface area and $\alb$ is the effective, planetary bond albedo.

It is important to note that, because $\tequ$ is derived from global energetic considerations, it is not necessarily close to the mean surface temperature, $\tsurf$, and is often not. This is for two main reasons:
\begin{itemize}
\item[$\bullet$] The most widely recognized of these reasons is that, when an absorbing atmosphere is present, the emitted flux does not directly come from the ground. Because of this greenhouse effect, the surface temperature of an isolated atmospheric column can be \textit{higher} than its equilibrium temperature. This explains why Venus and Earth mean surface temperatures are 737\,K and 288\,K while their effective equilibrium temperatures are around 231\,K and 255\,K (assuming an albedo of 0.75 and 0.29).
\item[$\bullet$] The second reason is the non-linearity of the blackbody bolometric emission with temperature. Because this emissions scales as $T^4$, it can be demonstrated that the spatial average of the \textit{local} equilibrium temperature (\eq{loceqtemp}) is necessarily \textit{smaller} than the effective equilibrium temperature defined by \eq{eqtemp}\footnote{Similar conclusions can be reached concerning the \textit{temporal} average of the equilibrium temperature, which is smaller than the equilibrium temperature computed for the temporally averaged flux. Whether this behavior has an important impact on the physical temperature, however, depends on the ratio of the thermal equilibration timescale of the atmosphere and ground to the period of the irradiation variation. For the Earth, thanks to the oceans, the thermal inertia is large enough so that the mean flux seems to be the relevant quantity for both diurnal and seasonal variations.}.
Rigorously, this stems from the fact that $\ltequ$ is a concave function of the absorbed flux which, with the help of the Rogers-H\"{o}lder inequality \citep{Rog88,Hol89}, yields
\balign{\label{teqinequality}\frac{1}{S}\int \ltequ\d S\leqslant \left(\frac{1}{S}\int \ltequ^{4}\d S\right)^{1/4}\equiv\tequ.}
In other words, regions receiving twice as much flux do not need to be twice as hot to reach equilibrium and cold regions have a more important weight in the mean temperature.
\end{itemize}

To give an idea of how strong this second effect can be, let us consider the limiting case of an airless, synchronous planet in thermal equilibrium with a constant surface albedo, $\alb$. In this case, the local surface temperature is equal to the local equilibrium temperature and the \textit{mean} surface temperature is given by
\balign{
\tsurf&=\frac{1}{4\pi}\int_0^{2\pi}\d \phi\int_0^{1}\d \mus \left(\frac{(1-\alb)\,\Fs\,\mus}{\ssb}\right)^{1/4}\nonumber\\
&=\frac{1}{4\pi} \left(\frac{(1-\alb)\,\Fs}{\ssb}\right)^{1/4}\int_0^{2\pi}\d \phi\,\int_0^{1}\mus^{1/4}\d \mus.}
Performing the integrals, and using \eq{eqtemp}, we find
\balign{
\tsurf&=\frac{2\sqrt{2}}{5}\tequ\approx 0.57\,\tequ.}
While any energy redistribution by the atmosphere will tend to lower this difference\footnote{Note that if energy redistribution affects the mean surface temperature, it has no impact on the mean \textit{equilibrium} temperature. The 1/4 factor found in \eq{eqtemp} and the inequality in \eq{teqinequality} are exact and do not rely on any assumption about this redistribution or the thickness of the atmosphere. In fact, \eq{teqinequality} can be generalized to temperature means at any pressure levels in an atmosphere by $\frac{1}{S}\int T(p)\d S\leqslant \left(\frac{1}{S}\int T^{4}(p)\d S\right)^{1/4}$, although this does not have any implications for the fluxes at the given level. In particular the inequality $\tsurf\leqslant\tequ$ has no reason to hold when a sufficiently opaque atmosphere is present because of the greenhouse effect; Venus being a perfect example.
}, one needs to remember that even the mean surface temperature can be much lower than the effective equilibrium temperature.

These simple considerations show that equilibrium temperature can be misleading and should be used with great care. Indeed, if 1D models can take into account the greenhouse effect, they cannot predict energy redistribution on the night side. Consequently, not only can they not predict temperatures on the nightside, but they also predict fairly inaccurate mean temperatures. Thus, while equilibrium temperature remains a handy tool to compare the temperature regime from one planet to another, it is not accurate enough to draw conclusions on the climate of a given object.

\subsection{Climate on a dry synchronous planet}
\label{sec:dry_runs}

\begin{figure*}[htbp]
\begin{center}
 \subfigure{ \includegraphics[scale=.73,trim = 0cm 1.cm .65cm 0cm, clip]{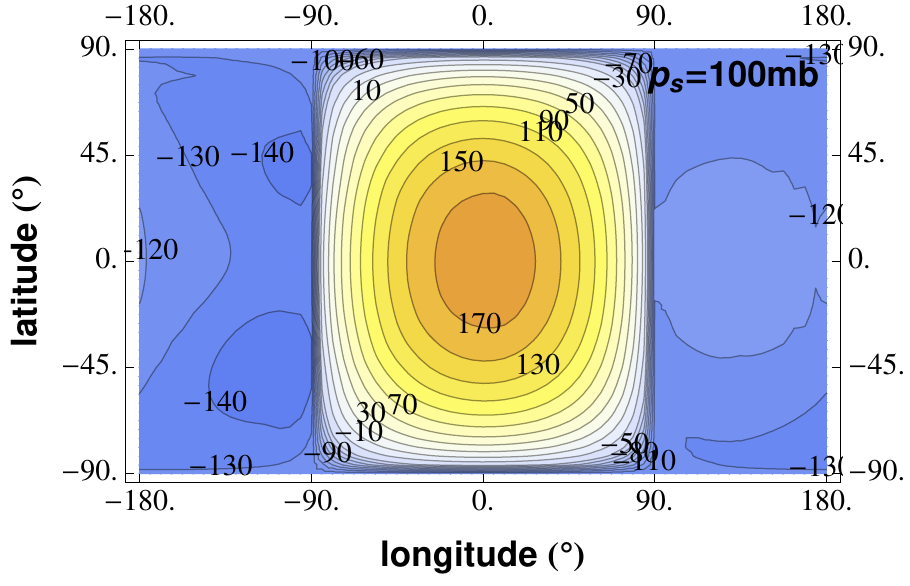} }
 \subfigure{ \includegraphics[scale=.73,trim = 1.2cm 1.cm .65cm 0cm, clip]{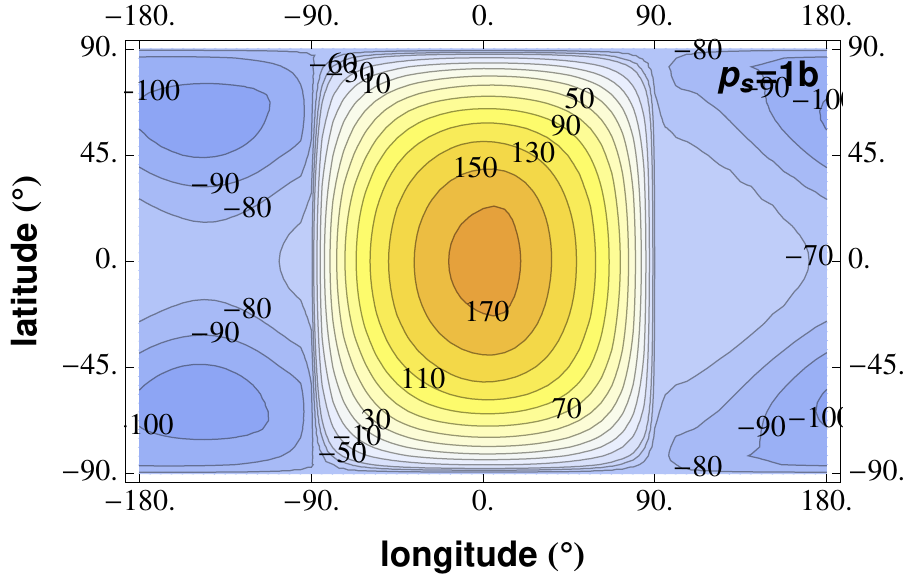} }
 \subfigure{ \includegraphics[scale=.73,trim = 1.2cm 1.cm 0cm 0cm, clip]{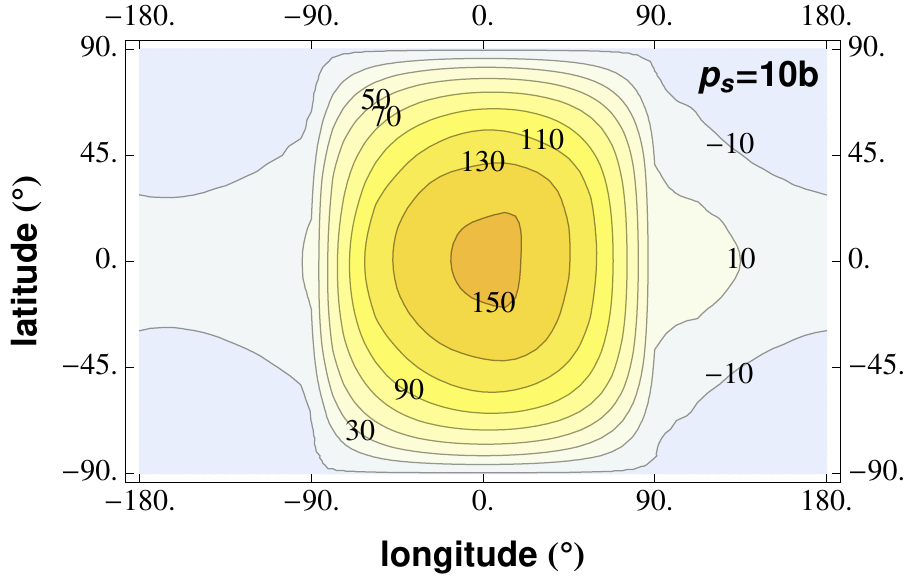} }\\
 \subfigure{ \includegraphics[scale=.73,trim = 0cm 0cm .65cm .3cm, clip]{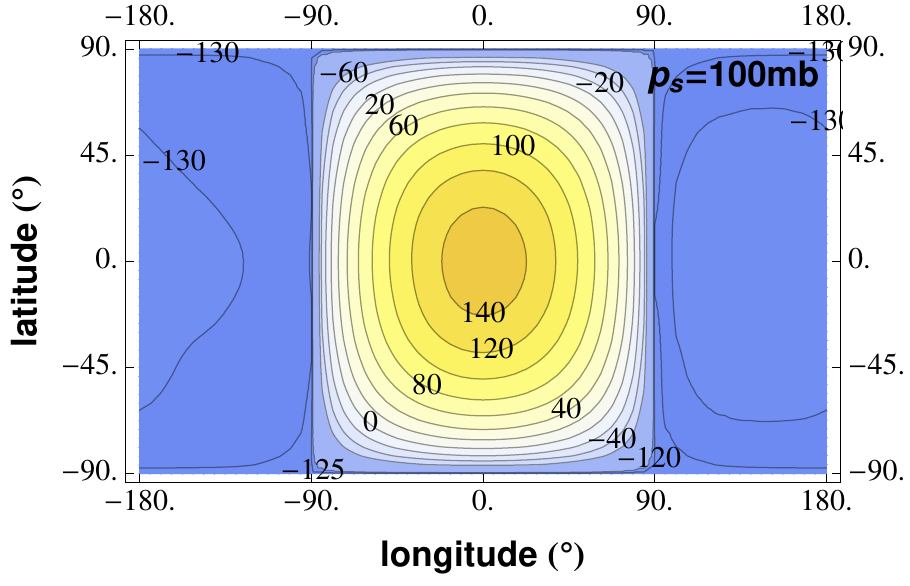} }
 \subfigure{ \includegraphics[scale=.73,trim = 1.2cm 0cm .65cm .3cm, clip]{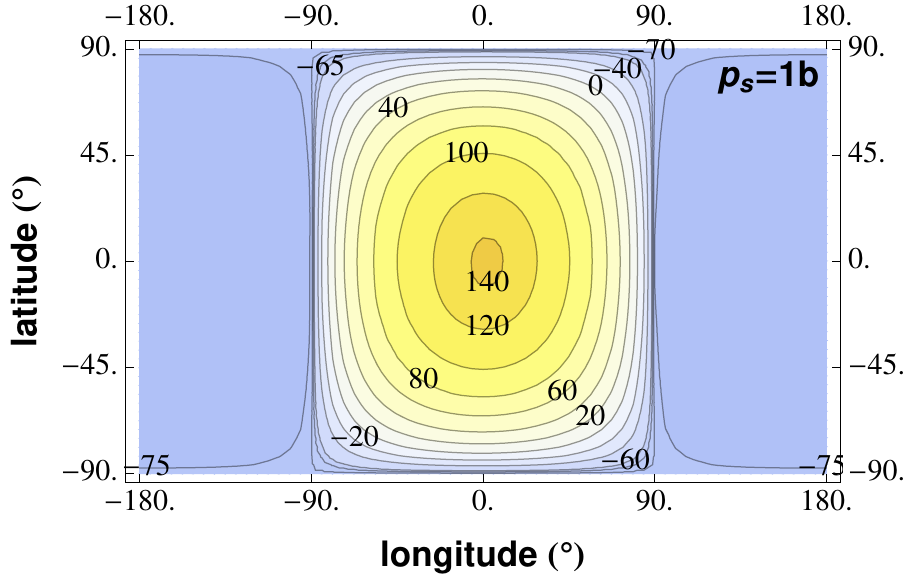}}
 \subfigure{ \includegraphics[scale=.73,trim = 1.2cm 0cm 0cm .3cm, clip]{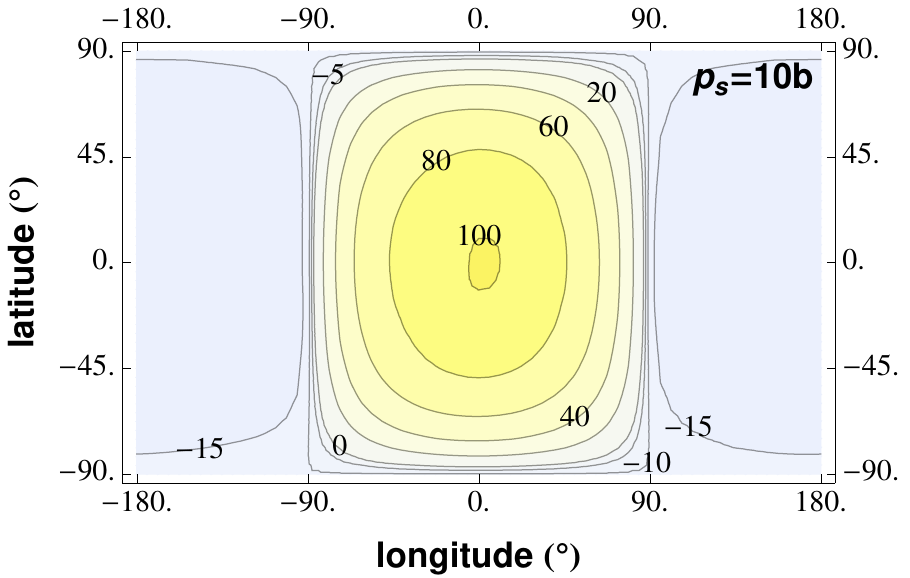}} \\
\end{center}
\caption{Surface temperature maps (in $^{\circ}$C) for the dry \gl581 (top row) and \hd85 (bottom row) runs for a total pressure ($\ps$) of 0.1, 1 and 10 bars (from left to right). Longitude 0$^\circ$ corresponds to substellar point.
}
\label{fig:surf_temp_map}
\end{figure*}

We now present a first set of 3D simulations of \gl581 and \hd85 that are modeled as completely dry planets. The goal of these idealized simulations is to gain insight into the various circulation patterns arising on synchronously rotating terrestrial planets and how the circulation regime depends on the atmospheric mass (mean surface pressure) and planetary rotation rate.

\subsubsection{Surface temperatures}

These dry runs have been performed using the parameters from \tab{tab:params}, without a water cycle. For various surface pressures, the model has been initialized from a rest state with uniform surface temperature and integrated until statistical equilibrium is reached.

The temperature maps for three sample surface pressures for the two planets are shown in \fig{fig:surf_temp_map}. The most apparent feature is the expected dichotomy between the sunlit (between -90$^\circ$ and 90$^\circ$ longitude) and dark hemispheres. 
A surprising outcome of these simulations is the great magnitude of the temperature difference between the substellar point and the night side compared to the one in synchronous Earth simulations of \citet{Jos03} and \citet{MS10} or to the equator-pole temperature difference on Earth. To understand these differences, we have performed an additional suite of simulations of fast rotating or tidally synchronized Earth with our full radiative transfer or with a gray gas approximation (as in \citealt{MS10}) and with or without water cycle. The model recovers both Earth present climatology and results similar to \citet{MS10}.

Thanks to this suite of simulations, the causes of this important temperature contrast are found to be (more or less by order of importance):
\begin{enumerate}
\item Latent cooling/heating. When water is abundantly present, the temperature at the substellar point is controlled mostly thermodynamically and is much lower than in the dry case. Latent energy transported by the atmosphere to the night side also contributes significantly to the warming of the latter. 
\item Non gray opacities. In our case dry case, greenhouse is solely due to the 15$\mu$m band of \co2 (especially on the night side). Thus, using a gray opacity of the background atmosphere with an optical depth near unity, as done in \citet{MS10}, considerably underestimates the cooling of the surface and thus overestimates the night side temperature.
\item A different planet. When its bulk density is constant, both the size and the surface gravity increase with the planet's size. Apart from the dynamical effect due to the variation of the Rossby number (extensively discussed below), note that for a given surface pressure ($\ps$), both the mass of the atmosphere and its radiative timescale ($\trad$) decrease with gravity (and with incoming flux for the latter) and that the advective timescale ($\tadv$) increases with the radius. For a given surface pressure the energy transport, thought to scale with the atmospheric mass and $\trad/\tadv$, thus tends to be less efficient on a more massive and more irradiated object.
\end{enumerate}

\begin{figure}[tbp]
\begin{center}
 \resizebox{.8\hsize}{!}{\includegraphics{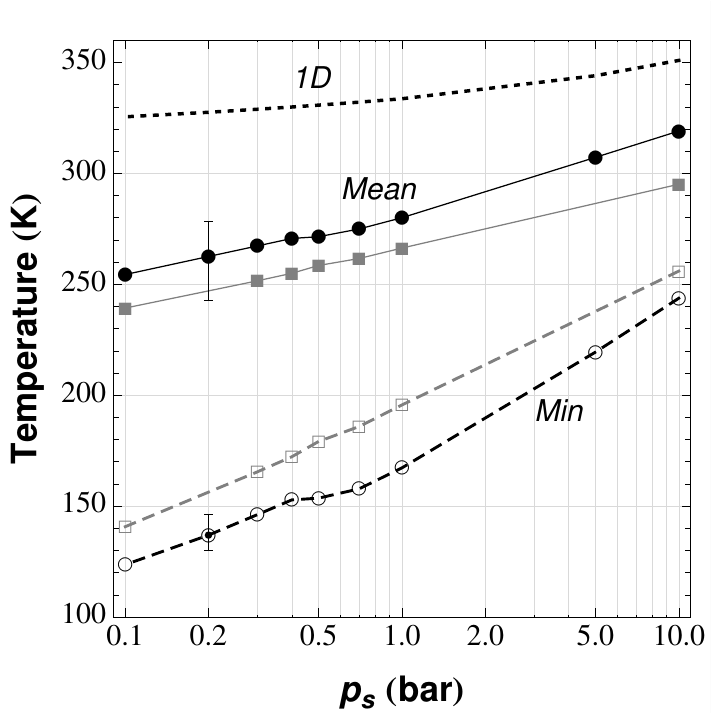}}
 \end{center}
\caption{Globally averaged (solid, filled) and minimal (dashed, empty) temperatures as a function of surface pressure for \gl581 (black circles) and \hd85 (gray squares). For comparison, the surface temperature of the 1D model for \gl581 is also shown (dotted curve). The error bars on the 200\,mbar case show the temperatures reached by varying the surface albedo between 0.1 and 0.5. }
\label{fig:Ts_vs_ps}
\end{figure}

Indeed, as expected, the temperature difference decreases with increasing total pressure, because the strength of the advection of sensible heat to the night side increases with atmospheric mass.
Both because of this more efficient redistribution and of the greenhouse effect of the atmosphere, the globally averaged mean surface temperature increases with surface pressure, as shown in \fig{fig:Ts_vs_ps}.

For comparison, we also computed the surface temperature yielded by the 1D version of our model. In this case, the atmospheric column is illuminated with a constant flux equal to the globally averaged impinging flux (1/4 of the substellar value) and with a stellar zenith angle of 60$^\circ$. Apart from this prescription for the flux and the fact that large-scale dynamics is turned off, our 1D code is exactly the same as our 3D one. As can be seen in \fig{fig:Ts_vs_ps}, even with a 10\,bar atmosphere, the 1D model is still 30\,K hotter than the globally averaged 3D temperature. This difference increases up to 70\,K for the 100\,mb case because redistribution is less efficient (see \sect{sec:analytical}). Note that a surface albedo of 0.3 is assumed here, similar to that in desert regions on Earth. The sensitivity of the surface temperature when albedo is varied between 0.1 and 0.5 is shown in \fig{fig:Ts_vs_ps}.

A troubling, and quite counterintuitive outcome of these simulations is that, as shown in \fig{fig:Ts_vs_ps}, although \gl581 receives more flux and has a higher mean surface temperature than \hd85, its minimal temperatures are always lower. As can be seen in \fig{fig:surf_temp_map}, the models for \gl581 present very cold regions located at mid latitudes and west from the western terminator (also called morning limb), whereas the \hd85 models have a rather uniform (and higher in average) temperature over the whole nightside.

\subsubsection{Circulation regime}

\begin{figure*}[htbp]
\begin{center}
 \subfigure{ \includegraphics[scale=.73,trim = 0cm 1.cm .65cm 0cm, clip]{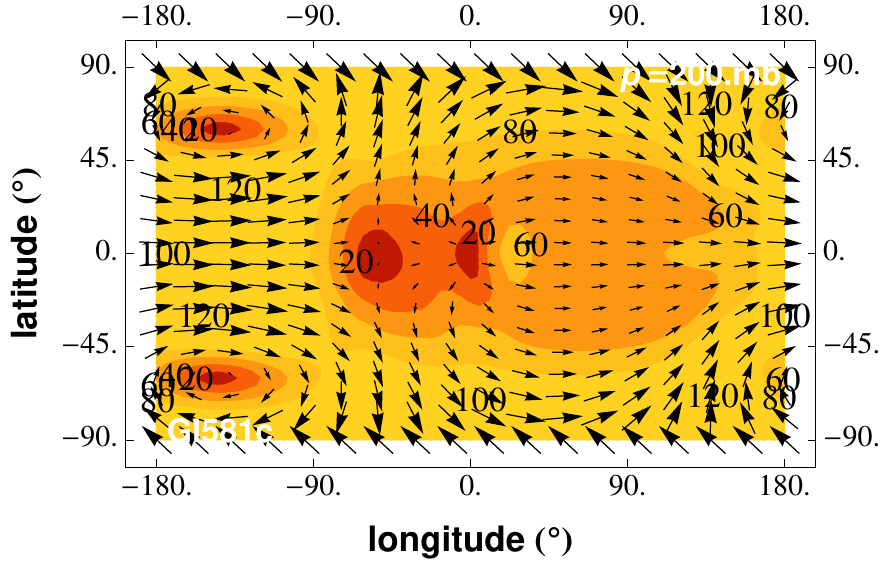} }
 \subfigure{ \includegraphics[scale=.73,trim = 1.2cm 1.cm .65cm 0cm, clip]{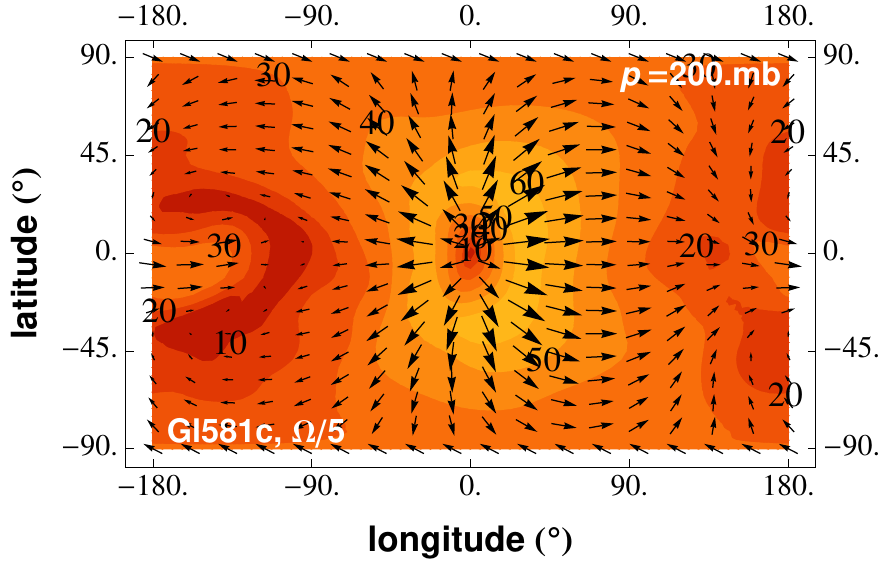} }
 \subfigure{ \includegraphics[scale=.73,trim = 1.2cm 1.cm 0cm 0cm, clip]{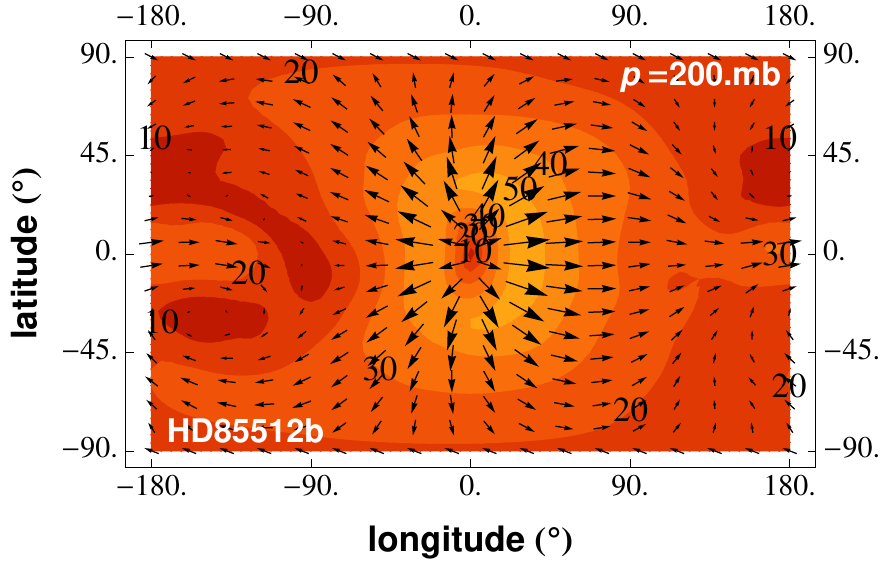} }\\
 \subfigure{ \includegraphics[scale=.73,trim = 0cm 0cm .65cm .3cm, clip]{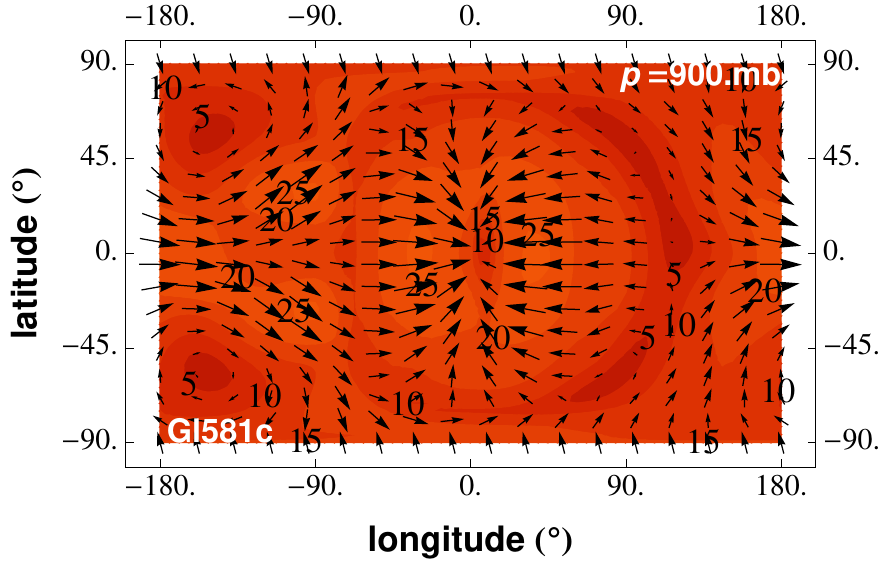} }
 \subfigure{ \includegraphics[scale=.73,trim = 1.2cm 0cm .65cm .3cm, clip]{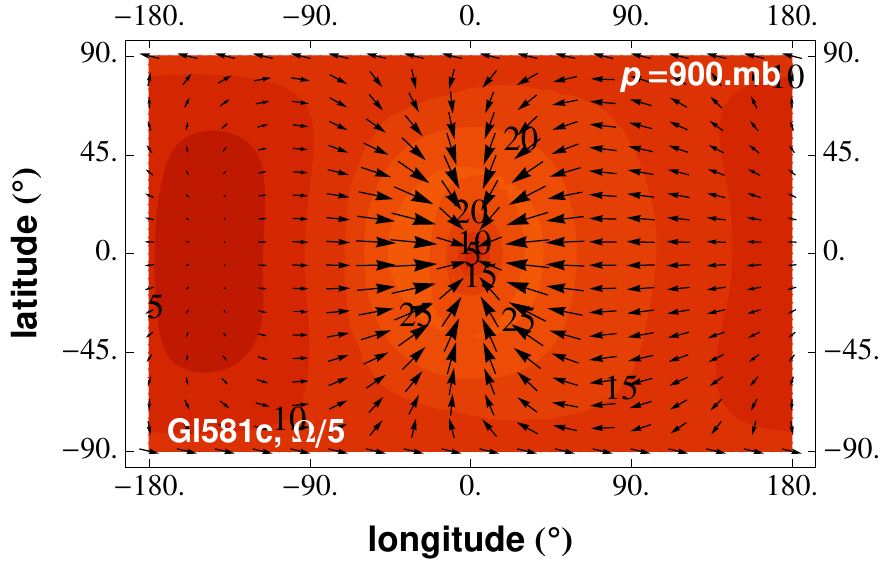}}
 \subfigure{ \includegraphics[scale=.73,trim = 1.2cm 0cm 0cm .3cm, clip]{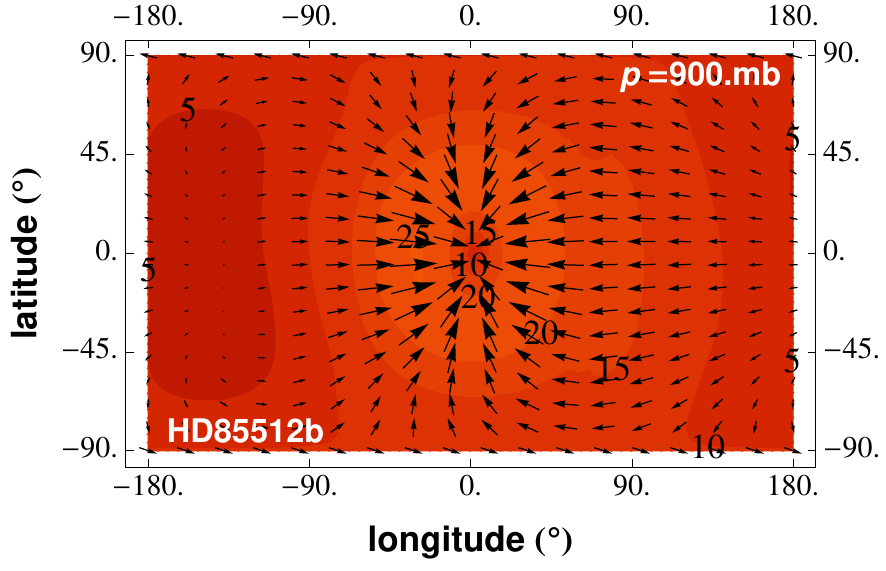}} \\
\end{center}
\caption{Wind maps (arrows) at the 200\,mb pressure level (top panel) and near the surface (900\,mb pressure level; bottom panel) for \gl581, \gl581 with a slower rotation (see text), and \hd85 (from left to right respectively). The mean surface pressure is 1\,b for all the models. Color scale (same for all panels) and numbers show the wind speed in m/s. Air is converging toward the substellar point at low altitudes in all models. However, whereas the standard \gl581 case clearly exhibits eastward jets (super-rotation) at higher altitudes, both slowly rotating cases tend to have stellar/anti stellar winds at the same level.
}
\label{fig:wind_map}
\end{figure*}

Further investigations show that this difference is a signature of the distinct circulation regimes arising on the two planets. Indeed, as shown in \fig{fig:wind_map}, our simulations of the atmosphere of \gl581 develop eastward jets (super-rotation) as found in many GCM of tidally locked exoplanets \citep{SMC08,TC10,HV11,HFP11,SCM11,SP11,SFN13}. Conversely, the atmosphere of \hd85 seems to predominantly settle into a regime of stellar/anti stellar circulation where high altitude winds blow from the dayside to nightside in a symmetric pattern with respect to the substellar point (upper right panel of \fig{fig:wind_map}; similar to the slowly rotating case of \citealt{MS10}).

To understand these distinct circulation regimes, one needs to understand the mechanism responsible for the onset of super-rotation on tidally locked exoplanets. \citet{SP11} showed that the strong longitudinal variations in radiative heating causes the formation of standing, planetary-scale equatorial Rossby and Kelvin waves. By their interaction with the mean flow, these waves transport eastward momentum from the mid latitudes toward the equator, creating and maintaining an equatorial jet. As a result, they predict that the strength of the jet increases with the day/night temperature difference and that the latitudinal width of this jet scales as the equatorial Rossby deformation radius
\balign{\Lro\equiv\sqrt{\frac{\Nbv \,\hp}{\beta}},}
where $\Nbv$ is the \brunt frequency, $\hp$ is the pressure scale height of the atmosphere ($\Nbv \,\hp$ being the typical speed of gravity waves), and $\beta\equiv \partial f /\partial y$ is the latitudinal derivative of the Coriolis parameter, $\coriolis\equiv 2\, \op\,\sin \theta$, with $\op$ the planet rotation rate and $\theta$ the latitude ($\beta=2\,\op/\Rp$ at the equator). Roughly, the equatorial Rossby deformation radius is the typical length scale over which an internal (gravity) wave created at the equator is significantly affected by the Coriolis force. This allows us to define a dimensionless equatorial Rossby deformation length
\balign{\Ro\equiv\frac{\Lro}{\Rp}=\sqrt{\frac{\Nbv \,\hp}{2\,\op\,\Rp}}.}

\begin{figure*}[htbp]
\begin{center}
 \subfigure{ \includegraphics[scale=.9,trim = 0cm 0cm .0cm 0cm, clip]{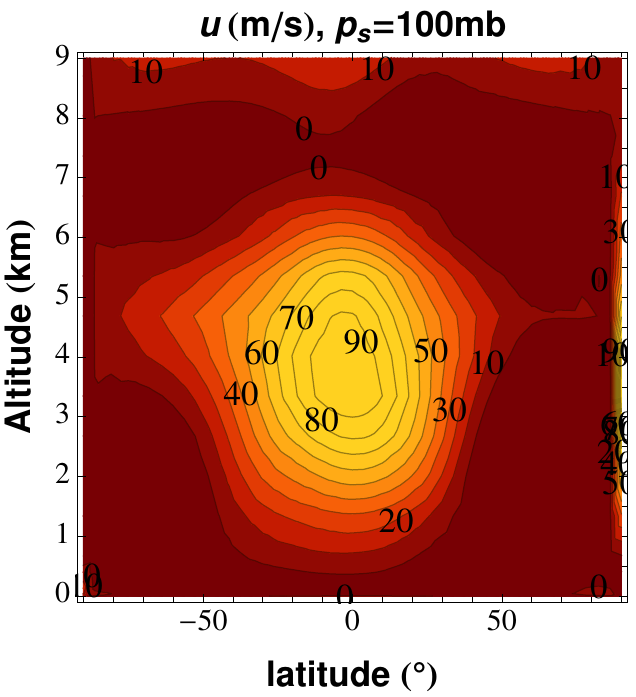} }
 \subfigure{ \includegraphics[scale=.9,trim = 0.8cm 0cm .0cm 0cm, clip]{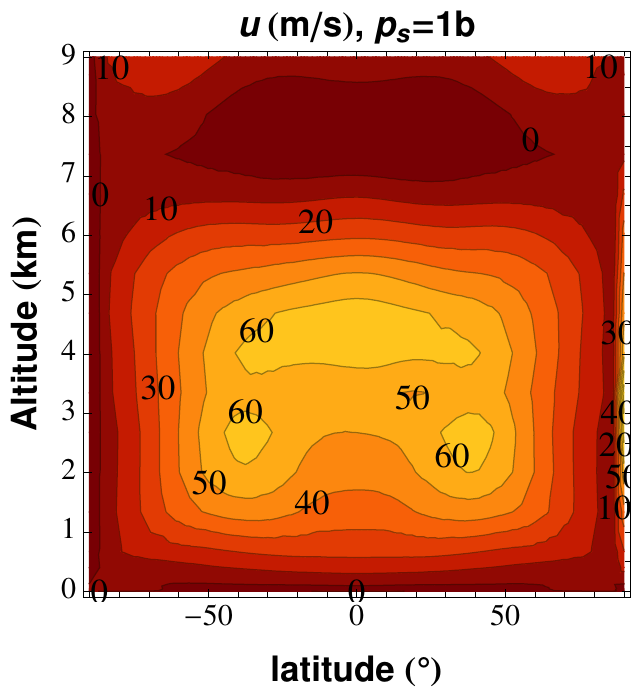} }
 \subfigure{ \includegraphics[scale=.9,trim = 0.8cm 0cm 0.cm 0cm, clip]{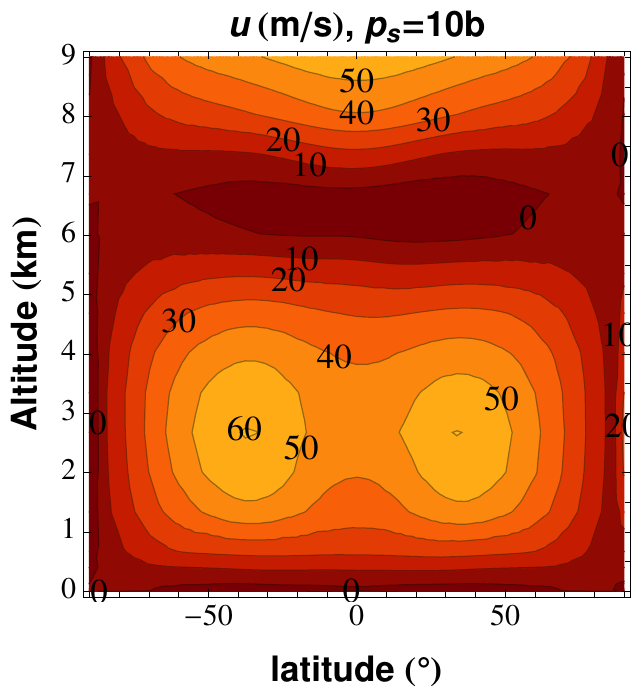}}
\end{center}
\caption{Zonally averaged zonal winds (in m/s) for the dry \gl581 runs for a total surface pressure of 0.1, 1 and 10 bars (from left to right).
}
\label{fig:zonal_wind_map}
\end{figure*}

This latter prediction is directly relevant to our case.
Indeed, although they have rather similar sizes and gravity, \gl581 rotates around its star in $\sim 13$\,Earth days (d) whereas \hd85, which orbits a brighter K star, has a $\sim 58$\,d orbital period. If spin orbit synchronization has occurred, the Coriolis force is thus expected to have a much weaker effect on the circulation of \hd85. To quantify this, let us evaluate $\Ro$ in our case. In a dry stably stratified atmosphere with a lapse rate $\d T/\d z$, the \brunt frequency is given by
\balign{\Nbv^2=\frac{\grav}{T}\left(\frac{\grav}{\cp}+\dd{T}{\,z}\right),}
$\cp$ being the heat capacity of the gas at constant pressure. The pressure scale height is given by
\balign{\hp\equiv\frac{\kB T}{m_\mathrm{a}\, \grav},}
where $\kB$ is the Boltzmann constant and $m_\mathrm{a}$ is the mean molecular weight (in kg) of the air. Assuming an isothermal atmosphere, this yields
\balign{\Ro=\sqrt{\frac{\kB}{m_\mathrm{a}\,\cp^{1/2} }\frac{T^{1/2}}{2 \,\op\,\Rp}}.}

Assuming a \N2 dominated atmosphere, using the equilibrium temperature as a proxy for $T$ and disregarding the latitude factor, one finds $\Ro=1.1$ for \gl581 and 2.5 for \hd85. In our slowly rotating case, the planet is thus too small to contain a planetary Rossby wave and the mechanism described above is too weak to maintain super-rotation. This also explains why the jet is planet-wide in the more rapidly rotating case of \gl581. To confirm this analysis, we have run a simulation of \gl581 where we have, everything else being equal, reduced the rotation rate by a factor of five so that the $\Ro$ is the same than for \hd85 ($\op=\Omega_\mathrm{orb}/5$ case; middle column of \fig{fig:wind_map}). As expected from the dimensional analysis, in this slowly rotating case, no jets appear and we recover a stellar/antistellar circulation. Another argument in favor of this analysis is the fact the location and shape of the cold regions and wind vortices present in the \gl581 case (see \figs{fig:surf_temp_map}{fig:wind_map}) are strongly reminiscent of Rossby-wave gyres 
seen in the Matsuno-Gill standing wave pattern \citep{Mat66,Gil80,SP11}.

Another way to look at this is to consider a multi-way force balance \citep{SFN13}. A way to tilt eddy velocities northwest-southeast in the northern hemisphere and southwest-northeast in the southern hemisphere, driving super-rotation, is to have a force balance between pressure-gradient ($\nabla_p \Phi\equiv \kB \Delta T \Delta \ln p /(m_\mathrm{a}\Rp)$, where $\Delta T$ is the horizontal day-night temperature difference and $\Delta \ln p$ is logarithmic pressure range over which this temperature difference is maintained), Coriolis ($\approx f\,u$ where $u$ is the wind speed), and advection ($\approx u^2/\Rp$) acceleration (see \citet{SFN13} for a discussion on the effect of drag). In this case one should have \balign{\nabla_p \Phi\sim f\,u\sim \frac{u^2}{\Rp},} which leads to
$u\sim f\, \Rp$ and eventually $f\sim \sqrt{\nabla_p \Phi/\Rp}$. 

If $f\ll \sqrt{\nabla_p \Phi/\Rp}$ this three-way force balance should reduce to a two way force balance between advection and pressure-gradient accelerations which leads to symmetric day-night flow. A numerical estimate with the same parameters as above with $\Delta T=100\,$K and $\Delta \ln p=1$ yields a critical rotation period of about 10\,d, close to \gl581 rotation period but much smaller than the one of \hd85. This confirms that \hd85 should have a much more pronounced day-night circulation pattern than \gl581, but it also tells us that Coriolis force is already quite weak in the force balance of \gl581 explaining the significant meridional wind speeds seen in \fig{fig:wind_map}.

There are a few general differences with previous close-in giant planet 3D simulations, however. First, in all our simulations, low altitude winds converge toward the substellar point. This results from the strong depression caused there by a net mass flux horizontal divergence high in the atmosphere driving a large-scale upward motion. Second, even at levels where eastward winds are the strongest (in the simulations with super-rotation), the wind speeds exhibit a minimum (possibly negative) along the equator, westward from the substellar point. At a fixed altitude, this results from a pressure maximum occurring there because of the higher temperature, and consequently higher pressure scale height.

\begin{figure}[htbp]
\begin{center}
 \resizebox{.8\hsize}{!}{\includegraphics{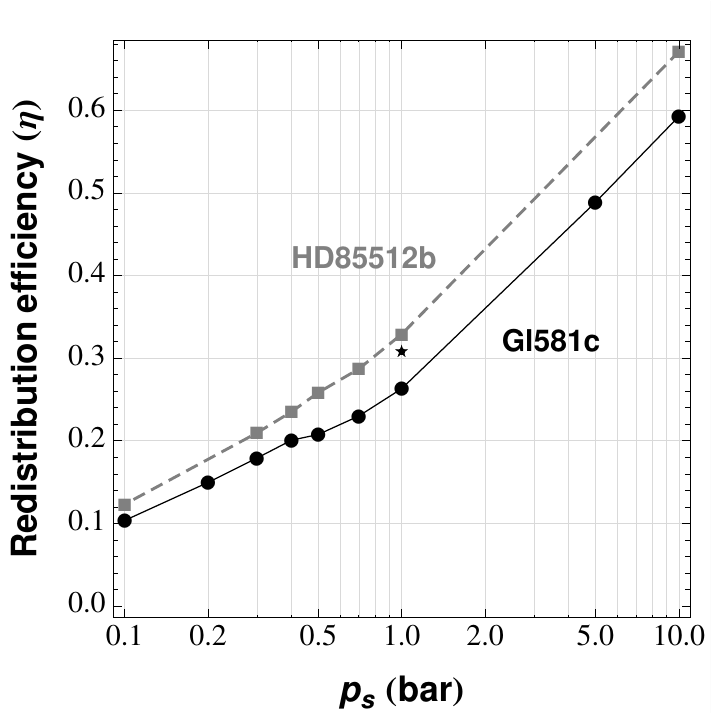}}
 \end{center}
\caption{Redistribution efficiency $\redist$ (ratio of night to dayside outgoing thermal emission) as a function of atmospheric total pressure ($\ps$) for \gl581 (black circles) and \hd85 (gray squares). $\redist=0$ implies no energy redistribution at all whereas an homogeneous thermal emission would yield $\redist=1$. The star symbol shows the redistribution efficiency of the $\op=\Omega_\mathrm{orb}/5$ \gl581 case.}
\label{fig:redistrib}
\end{figure}

Another feature visible in \fig{fig:zonal_wind_map} is the dependence of the strength of the zonal jet to the atmospheric mean surface pressure (lower pressure causing faster winds). A detailed analysis is out of the scope of this study. Nevertheless, these simulations seem to confirm the fact that zonal wind speeds increase with forcing amplitude \citep{SP11}. Indeed, an increasing pressure entails a decreasing thermal contrasts between the two hemispheres and a smaller forcing, hence the lower wind speeds.

Surprisingly, the presence of a strong equatorial jet lowers the heat redistribution efficiency by the atmosphere. By focusing the flow around the equator, the Coriolis force prevents an efficient redistribution to the midlatitudes on the night side.
To quantify in a simple way this redistribution, we computed the ratio $\redist$ of the mean outgoing thermal flux that is emitted by the night side over the mean outgoing thermal flux from the day side (see \fig{fig:redistrib}). With this definition, no redistribution should yield $\redist=0$ whereas a very homogeneous emission would produce $\redist=1$. As expected, the redistribution increases with total pressure. At constant pressure, redistribution seems to be controlled by the equatorial Rossby number and is more important in the slowly rotating cases (\hd85 and $\op=\Omega_\mathrm{orb}/5$ \gl581 case).

Further idealized simulations are needed to fully understand the mechanisms controlling both the circulation pattern and the energy redistribution on slowly rotating, synchronous terrestrial planets.

\subsubsection{Thermal profile}

\begin{figure}[htbp]
\floatbox[{\capbeside\thisfloatsetup{capbesideposition={right,bottom},capbesidewidth=3.3cm}}]{figure}[\FBwidth]
{\caption{Temperature profiles of the 1\,bar case for \gl581 (Grey curves). The solid black curve stands for the spatially averaged profile of the 3D simulation. For comparison, the dashed curve represents the temperature profile obtained with the global 1D model.}
\label{fig:temp_profil}}
{\includegraphics[scale=1.]{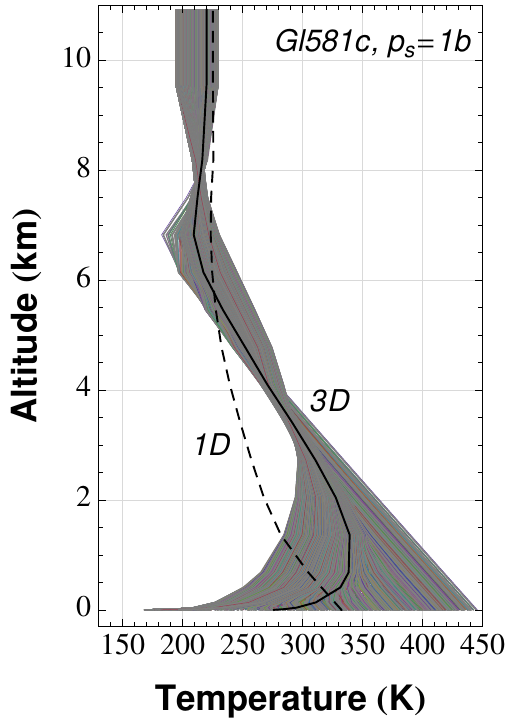}}
\end{figure}

Finally, in \fig{fig:temp_profil}, we present typical temperature profiles obtained from our simulations (here in the 1\,bar case, but profiles obtained for different pressures are qualitatively similar). For comparison, \fig{fig:temp_profil} also shows the profile obtained from the corresponding global 1D simulation (dashed curve), and the spatially averaged profile from our 3D run (solid curve).


One can directly see that, in the lower troposphere, temperature decreases with altitude on the day side following an adiabatic profile and increases on the night side. This temperature inversion on the night side is caused by the redistribution of heat, which is most efficient near 2-4\,km above the surface where winds are not damped by the turbulent boundary layer. The resulting mean temperature profile cannot be reproduced by the 1D model, which is to hot near the surface and too cold by more than 50\,K in the upper troposphere. The discrepancy tends to decrease in the upper stratosphere.

\section{Bistable climate: runaway vs cold trapping.}
\label{sec:coldtrap}

As inferred from 1D models (e.g. \citealt{KWR93,SKL07}) and confirmed by our 3D simulations (not shown), if water were present in sufficient amount everywhere on the surface (aquaplanet case), both our prototype planets (\gl581 and \hd85) would enter a runaway greenhouse state \citep{KPA84}. The large flux that they receive thus prevents the existence of a deep global ocean at their surface.

However, as visible in \fig{fig:surf_temp_map}, because of the rather inefficient redistribution, a large fraction of the surface present moderate temperatures for which either liquid water or ice would be thermodynamically stable at the surface. In the 10b case, the whole night side exhibits mean temperatures that are similar to the annual mean in Alaska. It is thus very tempting to declare these planets habitable because there always exist a fraction of the surface where liquid water could flow. While tempting, one should not jump to this conclusion because of the two following points.
\begin{itemize}
\item [$\bullet$] Even if water is thermodynamically stable, it is always evaporating. Consequently, the water vapor is transported by the atmosphere until it condenses and precipitates (rains or snows) onto the ground in cold regions. This mechanism tends to dry warmer regions and to accumulate condensed water in cold traps. If these cold traps are permanent, or if no mechanism is physically removing the water from them - e.g. ice flow due to the gravitational flattening of an ice cap, equilibration of sea level (see \sect{sec:icecap}) - this transport is irreversible.

\item [$\bullet$] Because it is a strong greenhouse gas, water vapor has an important positive feedback. The presence of a small amount of water can dramatically increase surface temperatures and even trigger the runaway greenhouse instability.
\end{itemize}

Therefore, even if useful, "dry" simulations cannot be used to assess the climate of a planet where water is potentially present, and the distribution of the latter\footnote{At low stellar incoming fluxes, the presence of water also entails a positive feedback through the ice albedo feedback so that this argument still holds (see e.g. \citealt{Pie10})}. Hence, in this section we present a set of simulations taking into account the full water cycle but with a limited water inventory.

We will show that, as already suggested by \citet{AAS11}, climate regimes on heavily irradiated land planets result from the competition of the two aforementioned processes and lead to the existence of two stable equilibrium states. A "collapsed" state where almost all the water is trapped in permanent ice caps on the night side or near the poles, and a runaway greenhouse state where all the water is in the atmosphere. Furthermore, we show that the collapsed state can exist at much higher fluxes than inferred by \citet{AAS11}, and that there is no precise flux limit, as it can depend on both the atmosphere mass and water amount.

\subsection{Synchronous case}

\begin{figure}[htbp]
\begin{center}
 \resizebox{1.\hsize}{!}{\includegraphics[trim = 2cm 1.8cm 1.6cm 2cm, clip]{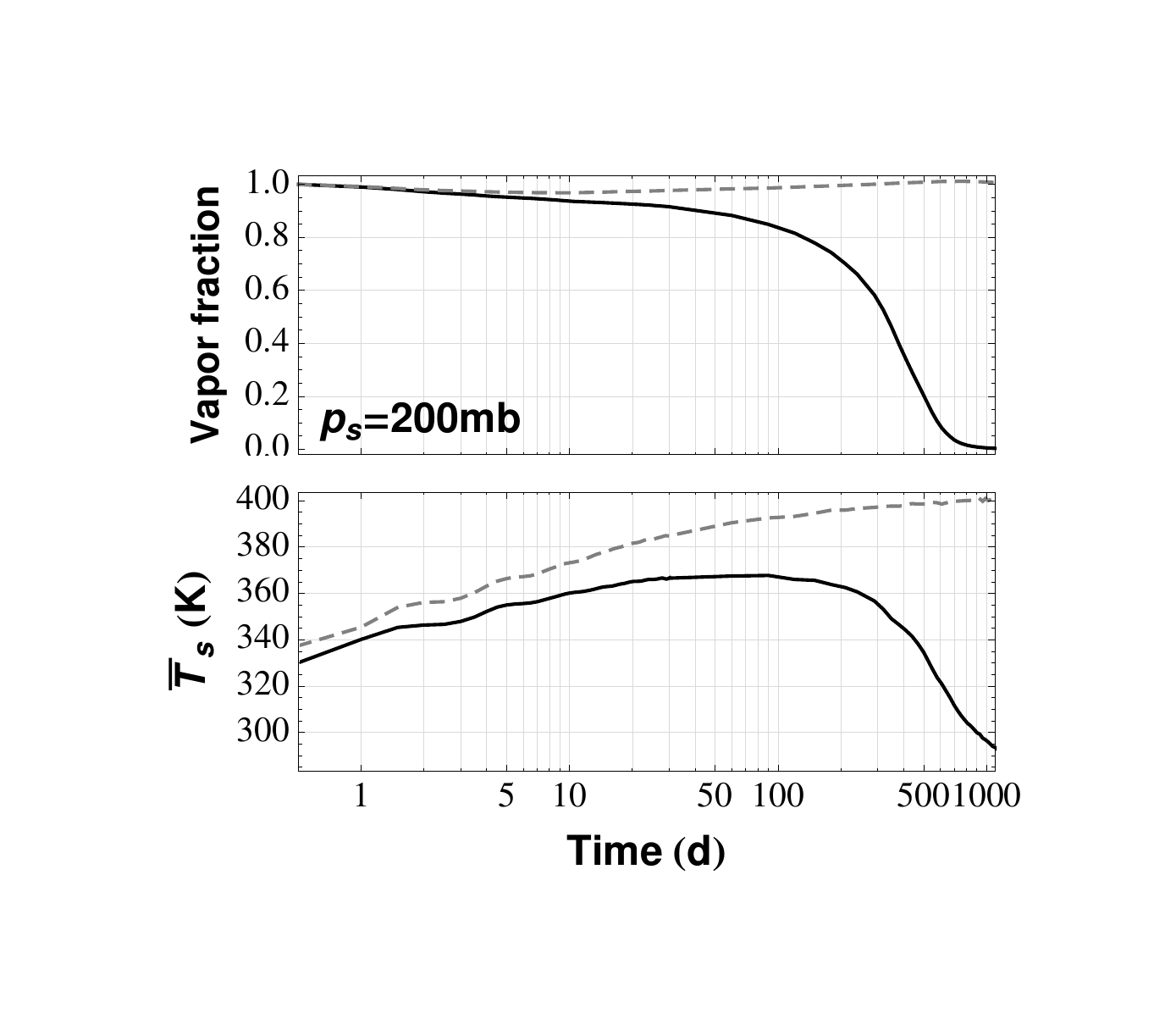}}
 \end{center}
\caption{Time evolution of the fraction of water vapor in the atmosphere over the total water amount and of the mean surface temperature for \gl581 with a 200\,mbar background atmosphere. The black solid curve represents a model initialized with a water vapor column of 150\,kg/m$^2$ and the gray dashed one with 250\,kg/m$^2$. After a short transitional period, either the water collapses on the ground in a few hundred days (solid curve) or the atmosphere reaches a runaway greenhouse state (dashed curve).}
\label{fig:vapstart}
\end{figure}

Our numerical experiment is as follows. For our two planets and for various pressures, we start from the equilibrium state of the corresponding "dry" run described in \sect{sec:dry_runs} to which a given amount of water vapor is added. To do so, a constant mass mixing ratio of water vapor, $\qvap$, is prescribed everywhere in the atmosphere, and surface pressure is properly rescaled to take into account the additional mass. Two typical examples of the temporal evolution of the surface temperature and water vapor content in these runs is presented in \fig{fig:vapstart}.

While idealized, this numerical experiment has the advantage of not introducing additional parameter to characterize the initial repartition of water. It also provides us with an upper limit on the impact of a sudden water vapor release as vapor is present in the upper atmosphere where its greenhouse effect is maximum. From another point of view, this experiment roughly mimics a planet-wide release of water by accretion of a large asteroid or planetesimal and can give us insight into the evolution of the water inventory after a sudden impact.

Because of the sudden addition of water vapor, the atmosphere is disturbed out of radiative balance. Water vapor significantly reduces the planetary albedo by absorbing a significant fraction of the stellar incoming radiation and it also reduces the outgoing thermal flux. The temperature thus rapidly increases until a quasi-static thermal equilibrium consistent with the amount of water vapor is reached in a few tens of days (see \fig{fig:vapstart}). At the same time, water vapor condenses, form clouds, and precipitate onto the ground near the coldest regions of the surface. As temperature increases, the amount of precipitation can decrease because the re-evaporation of the falling water droplets or ice grains becomes more efficient. The evaporation of surface water also becomes more efficient.

\begin{figure}[htbp]
\begin{center}
 \resizebox{1.\hsize}{!}{\includegraphics[trim = 1cm 0.2cm 1cm 0cm, clip]{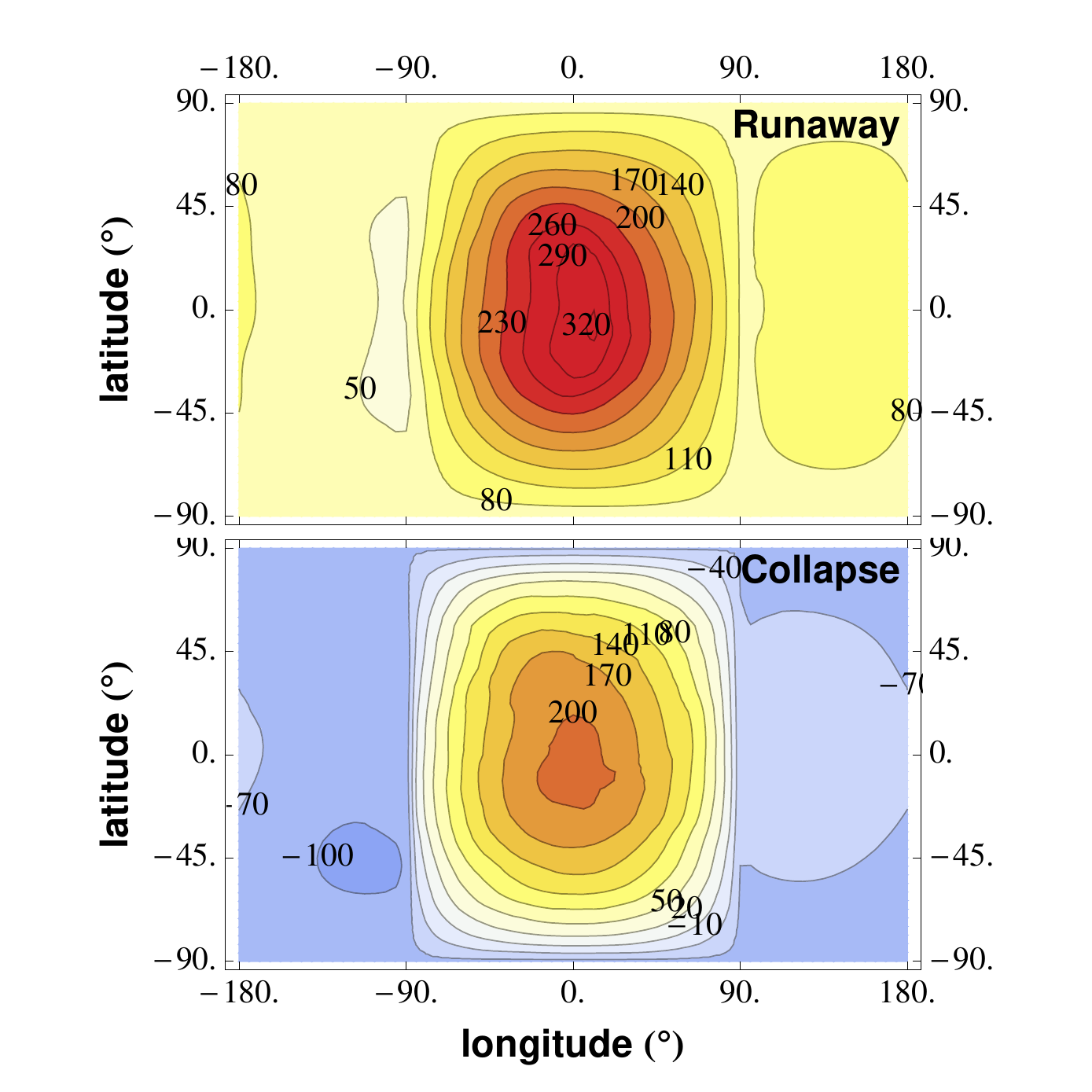}}
 \end{center}
\caption{Surface temperature map (in $^\circ$C) reached at equilibrium for the two cases depicted in \fig{fig:vapstart} (parameters of \gl581 with a 200\,mbar background atmosphere). The model shown in the top panel was initialized with a water vapor column of 250\,kg/m$^2$ and is in runaway greenhouse state. The bottom panel shows a collapsed state initialized with only 150\,kg/m$^2$ of water vapor.  }
\label{fig:TS_runaway_collapse}
\end{figure}

\begin{figure}[htbp]
\begin{center}
 \subfigure{ \includegraphics[scale=.95]{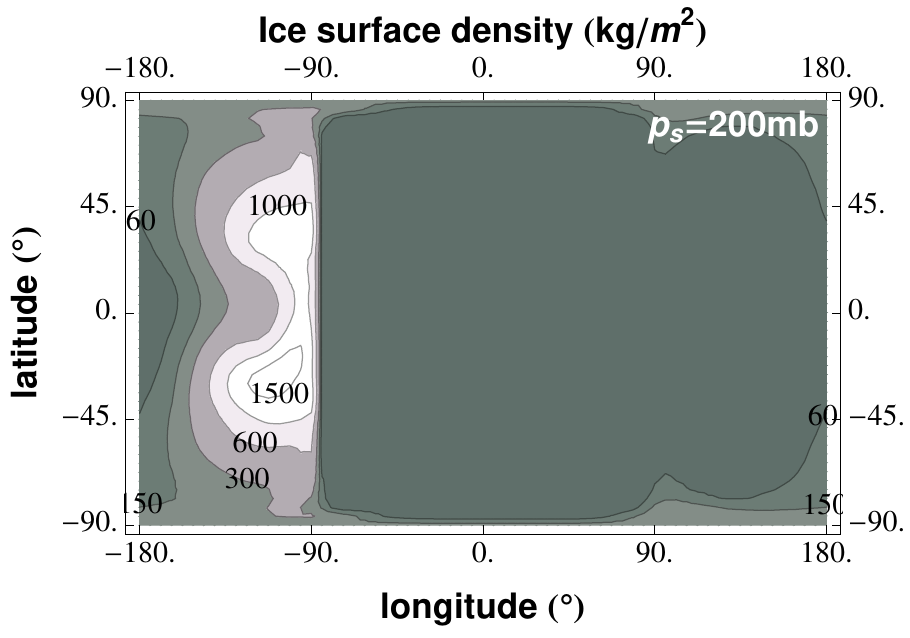} }
\end{center}
\caption{Ice surface density of the collapsed state of the 200\,mb case for \gl581 presented in \fig{fig:TS_runaway_collapse}. The average ice surface density is 150\,kg/m$^2$. Most of the ice has collapsed in the cold traps.
}
\label{fig:vapstart_ice}
\end{figure}

Then, the system can follow two different paths. If the initial amount of water vapor and the associated greenhouse effect are large enough, evaporation will overcome precipitation and the atmosphere will stay in a stable hot equilibrium state where all the water is vaporized (with the exception of a few clouds; dashed curve in \fig{fig:vapstart} and top panel in \fig{fig:TS_runaway_collapse}). Because this state is maintained by the greenhouse effect of water vapor and that any amount of water added at the surface is eventually vaporized, this state is the usual runaway greenhouse state \citep{NHA92}. On the contrary, for smaller water vapor concentrations, the temperature increase is not large enough and the water vapor collapses in the cold traps (solid curve in \fig{fig:vapstart} and bottom panel in \fig{fig:TS_runaway_collapse}). The amount of water vapor decreases and the surface slowly cools down again until most of the water is present in ice caps in the colder regions of the surface. An example of the ice surface density of a collapsed state at the end of the simulation is presented in \fig{fig:vapstart_ice}. Because sublimation/condensation processes are constantly occurring, this ice surface density may continue to evolve, but on a much longer time scale that is difficult to model (see \citealt{WFM13} for a discussion). Then, water vapor is rather well mixed within all the atmosphere, and its mixing ratio is mainly controlled by the saturation mass mixing ratio just above the warmest region of the ice caps. Depending on the simulations, this mass mixing ratio varies between 10$^{-4}$-10$^{-5}$.

To investigate the factor determining the selection of one of these two regimes, we have run a large grid of simulations varying both the total surface pressure and the initial amount of water vapor. As the main effect of water vapor is its impact on the radiative budget, the quantity of interest is the total water vapor column mass (also called water vapor path)
\balign{
\mvap\equiv\int_{z=0}^{\infty} \qvap \,\rho\, \d z=\int^{\ps}_{p=0} \qvap \, \frac{\d p}{\grav},
}
which is linked to the spectral optical depth of water vapor ($\tauvap$) by the relation $\d \tauvap =\kapvap \qvap \rho \d z= \kapvap \d \mvap$.

\begin{figure}[htbp]
\begin{center}
 \resizebox{1.\hsize}{!}{\includegraphics{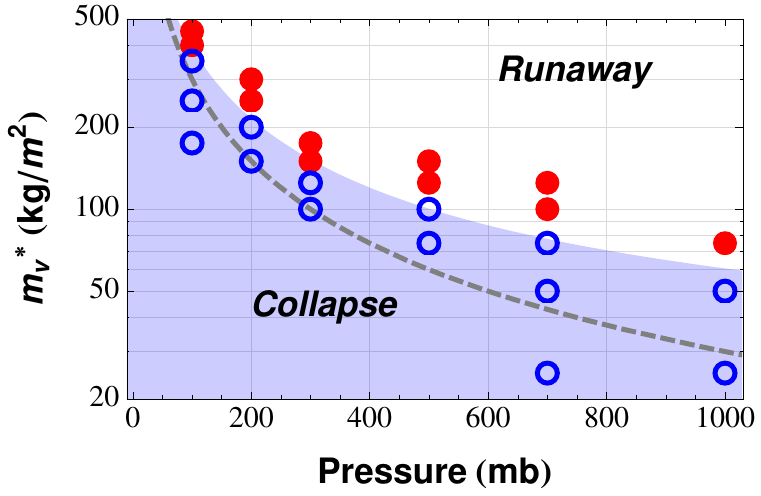}}
 \end{center}
\caption{Climate regime reached as function of the initial surface pressure and water vapor column mass ($\mvap$) for \gl581 runs. Red filled disks represent simulations that reached a runaway greenhouse states, and blue empty circles stand for simulation where water collapsed. The blue shaded region roughly depicts the region where collapse is observed. The gray dashed line represent the demarcation between runaway and collapse states for \hd85. At higher surface pressure, redistribution is large and cold traps are less efficient. Less water vapor is thus needed to trigger the runaway greenhouse instability.}
\label{fig:collapse_gl581}
\end{figure}

Results for the case of \gl581 are summarized in \fig{fig:collapse_gl581}. At a given pressure, a small initial amount of water vapor results in a collapsed case (blue empty circles). However, there always exists a critical initial water vapor path, function of the background pressure, above which the runaway greenhouse instability is triggered and develops (red disks). Because redistribution efficiency increases with atmospheric mass, cold trapping becomes weaker and the critical initial water vapor path decreases with background atmosphere surface pressure.

We also carried out simulations with a large amount of water (larger than the critical water vapor path discussed above), but distributed only at the surface in the cold traps. For a background atmosphere less massive than $\sim$1-5\,bar, these simulations are found to be stable against runaway. This demonstrates that, for a given total amount of water larger than the critical water vapor path, a "moist" bistability exists in the system. Owing to the inhomogeneous insolation, the 1D notion of critical flux above which runaway greenhouse becomes inevitable is no longer valid, highlighting the dynamical nature of the runaway greenhouse which needs a large enough initial amount of water vapor to be triggered.

When the insolation is decreased, a higher amount of water vapor is needed to trigger the runaway greenhouse instability, and this at every background pressure. As would be expected, planets receiving less flux are more prone to fall in a collapse state. However, unexpectedly, results for \hd85 do not follow this trend. At every background pressure, a lower amount of water vapor is needed to trigger the runaway greenhouse instability, as shown in \fig{fig:collapse_gl581}. This is due to the peculiar fact that, because of the higher redistribution efficiency caused by the lower rotation rate of the planet, cold traps are warmer (see \sect{sec:dry_runs} and \fig{fig:Ts_vs_ps} for details). As evaporation is mainly controlled by the saturation pressure of water above the cold traps, it is more efficient, destabilizing the climate.

In addition to the work of \citet{AAS11} which showed that runaway could be deferred to higher incoming fluxes on land planets, our simulations demonstrate that, even for a given flux, such heavily irradiated land planets can exhibit two stable climate in equilibrium. Furthermore, we have shown that the state reached by the atmosphere depends not only on the impinging flux, but also on the mass of the atmosphere, on the initial water distribution, and on the circulation regime (controlled by the forcing, the size of the planet and its rotation rate). This bistability is due to the dynamical competition between cold trapping and greenhouse effect.
An important consequence of this bistability is that the climate present on planets beyond the inner edge of the habitable zone will not only depend on their background atmosphere's mass and composition, but also on the history of the water delivery and escape, as discussed below.

\subsection{The 3:2 resonance}\label{sec:coldtrap3:2}

Because of the strong tidal interaction with their host star and of the small angular momentum contained in the planetary spin, rocky planets within or closer than the habitable zone of M and K stars are thought to be rapidly synchronized. Indeed, standard theory of equilibrium tides \citep{Dar1880,hut81,LCB10} predicts that a planet should synchronize on a timescale equal to
\balign{\tau_\mathrm{syn}=\frac{1}{3}r_\mathrm{g}^2 \frac{\Mp\, a^6}{G \Ms^2\Rp^3 \kp\,\dtp},}
where $r_\mathrm{g}$ is the dimensionless gyration radius ($r_\mathrm{g}^2=2/5$ for an homogeneous interior; \citealt{LLC11}), $G$ is the universal gravitational constant, $\kp$ is the tidal Love number of degree 2 and $\dtp$ is a time lag which characterizes the efficiency of the tidal dissipation into the planet's interior (the higher $\dtp$, the higher the dissipation). Other variables are defined in \tab{tab:params}. While the magnitude of the tidal dissipation in massive terrestrial planets remains largely unconstrained \citep{Han10,BRL11}, one can have a rough idea of the orders of magnitude involved by using the time lag derived for the Earth, $\kp\,\dtp=0.305\times629$\,s \citep{NL97}. This yields a synchronization timescale of 25\,000\,yr for \gl581 and 10\,Myr for \hd85, preventing these planets from maintaining an initial obliquity and fast rotation rate \citep{HLB11}.

As demonstrated by \citet{MBE12} in the case Gl\,581\,d, which is more eccentric and further away from its parent star than \gl581, tidal synchronization is not the only possible spin state attainable by the planet. As for Mercury, if the planet started from an initially rapidly rotating state, it could have been trapped in multiple spin orbit resonances during its quick tidal spin down because of its eccentric orbit. As \gl581 also has a non-negligible eccentricity of 0.07$\pm0.06$ \citep{FBD11}, it has almost a 50\% probability of being captured in 3:2 resonance (depending on parameter used) before reaching the synchronization (Makarov \& Efroimsky; private communication, 2012). Higher order resonances are too weak.

To assess the impact of a non-synchronized rotation, we performed the analysis described above in the 3:2 resonance case. The rotation rate of the planet is thus three half of the orbital mean motion and the solar day lasts two orbital periods. As expected, depending on the thermal inertia of the ground and on the length of the day, the night side is hotter than in the synchronized case. More importantly, there is no permanent cold trap at low latitudes. However, like for Mercury \citep{LFG13}, because of the low obliquity, a small area near the poles remains cold enough to play this role and ice can be deposited there. As the magnitude of the large scale transport of water vapor is lower towards the pole compared to the night side (zonal winds are stronger than meridional ones), the trapping is less efficient. As a result, the transition towards the runaway greenhouse state discussed in the previous section is triggered with a lower initial water vapor path. The timescale needed to condense all the water vapor near the poles is also much longer.


\section{Evolution of the water inventory}\label{sec:ice_limits}

How stable are the two equilibrium state described above in the long term? It is possible that the climate regime exhibited by an irradiated land planet changes during its life depending on the rate of water delivery and escape of the atmosphere and/or water: triggering runaway when water delivery (or water ice sublimation by impacts) is too large and collapsing water when either the background atmosphere or the water vapor mass decreases below a certain threshold. Let us briefly put some limits on the fluxes involved.

\subsection{Atmospheric escape}

\begin{figure}[htbp] 
 \center
\includegraphics[scale=1.]{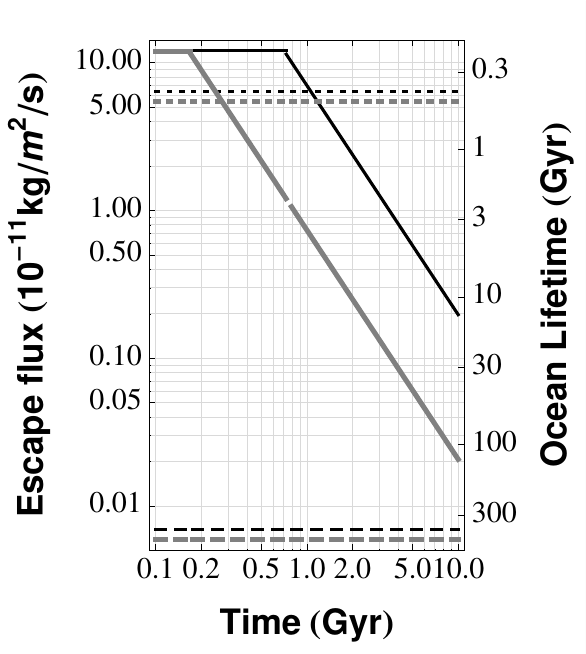}
 \caption{Energy limited escape flux as a function of stellar age ($\epsilon=0.15$; solid curve) for \gl581 (black) and \hd85 (gray). The dotted curve is the diffusion limit in the runaway state ($f_\mathrm{str}(\mathrm{H}_2)=0.1$ and $T_\mathrm{str}=300\,$K), and the dashed curve is an upper limit to the escape in the collapse state ($f_\mathrm{str}(\mathrm{H}_2)=10^{-4}$ and $T_\mathrm{str}=220\,$K). The time needed to evaporate an Earth ocean of water ($\approx1.4\times10^{21}$\,kg) is indicated on the right scale.}
 \label{fig:atmEscape}
\end{figure}

When water vapor is present in the upper atmosphere, XUV radiations can photo dissociate it and the light hydrogen atoms can escape\footnote{Of course, the energy available to photo dissociate water molecules themselves can also limit escape. Very simple calculations seem to suggest that this constraint is less stringent in our case, but this conclusion strongly depends on the assumed stellar spectral energy density in the XUV.}. An upper limit on this escape flux is provided by the energy limited flux (watson 1981) approximated by
\balign{F_\mathrm{el}= \epsilon \frac{\Rp F_\mathrm{XUV}}{G \Mp} \ \mathrm{(kg\,m^{-2}\,s^{-1})},}
where $F_\mathrm{XUV}$ is the energetic flux received by the planet and $\epsilon$ is the heating efficiency. The problem then lies in estimating the flux of energetic photons which varies with time (see \citealt{SKL07} for a discussion). Estimates of $F_\mathrm{el}$ are shown in \fig{fig:atmEscape} where the XUV flux parametrization of \citet{SRM10} and a heating efficiency of $\epsilon=0.15$ have been used\footnote{To give an idea of the impact on the water inventory, in \fig{fig:atmEscape} we estimated the lifetime of an Earth ocean by $\tau_\mathrm{ocean}=M_\mathrm{ocean}/(4\pi\Rp^2\,F_\mathrm{el})$, where $M_\mathrm{ocean}=1.4\times10^{21}$\,kg. This instantaneous estimate does not take into account the flux variation during the period considered.}. In some cases, this formula can also be used to estimate the escape of the background atmosphere itself.

In atmospheres where the amount of water vapor in the upper atmosphere is limited, the escape flux can be limited by the diffusion of water vapor in air. Following \citet{AAS11}, the diffusion-limited escape flux is written as
\balign{F_\mathrm{dl}= f_\mathrm{str}(\mathrm{H}_2) \,b_\mathrm{ia} \frac{\left(m_\mathrm{a}-m_\mathrm{i}\right)\,\grav } {\kB T_\mathrm{str}},}
where $f_\mathrm{str}(\mathrm{H}_2)$ is the mixing ratio of hydrogen in all forms in the stratosphere, $T_\mathrm{str}$ is the temperature there, and $m_\mathrm{a}$ and $m_\mathrm{i}$ are the particle mass of air and the species considered, and $b_\mathrm{ia}$ is the binary diffusion coefficient of i in air (for H$_2$ in air $b_\mathrm{ia}$= 1.9$\times10^{21}(T/300$\,K)$^{0.75}$m$^{-1}$s$^{-1}$).

In the collapsed state, the longitudinal and vertical cold traps are efficient. Temperature and water vapor mixing ratios are thus low in the stratosphere ($T_\mathrm{str}\approx 220K$ and $f_\mathrm{str}(\mathrm{H}_2)< 10^{-4}$; dashed curve in \fig{fig:atmEscape}). On the contrary, in the runaway state, the stratosphere temperature is much higher ($\approx\,300\,$K), and water is rather well mixed in the atmosphere. The mixing ratio can be directly calculated from the total water amount. To show an example, we plotted the case of $f_\mathrm{str}(\mathrm{H}_2)\approx 10^{-1}$ in \fig{fig:atmEscape}.

Of course, to have a comprehensive view of the problem, hydrodynamic and non-thermal (due to stellar winds) escape should also be accounted for. However, some conclusions can be made. First, it seems that escape of the background atmosphere could be efficient during the first few hundred million years to one billion year. This comes to support the possibility of a thin atmosphere. Second, as expected, escape of water vapor is always greater in the runaway state because of the higher temperature and humidity in the higher atmosphere \citep{KPA84}. Third, in the collapse state, cold traps limit the humidity of the upper atmosphere and escape is most likely diffusion-limited to a rather low value.

Interestingly enough, the sudden increase of water loss when runaway is triggered can provide a strong negative feedback on the water content of land planets beyond the classical inner edge of the habitable zone. Indeed, at least at early times where the XUV flux is strong enough, if a significant fraction of water is accreted suddenly and triggers the runaway, water escape will act to lower the water vapor amount until collapse ensues. This should happen when the atmospheric water vapor content is on the order of magnitude of, although lower than, the critical water column amount shown in \fig{fig:collapse_gl581}\footnote{In the numerical experiment shown in \fig{fig:collapse_gl581}, the critical water column amount is defined as the minimum water vapor amount to add to a \textit{cold initial state} to force its transition into a hot runaway state. This limit would probably be lower if we started with greater temperatures has cold trapping would be less efficient. We thus expect that the transition from runaway to collapse state occurs at smaller amounts of atmospheric water vapor}. Then, cold-trapped water has a much longer lifetime. This could particularly play a role during the early phases of accretion, where a runaway state is probable considering the important energy and water vapor flux due to the accretion of large planetesimals \citep{RQL07}, a possibly denser atmosphere, and a faster rotation. If the planet receives a moderate water amount, the water could be lost in a few tens to hundred million years, leaving time to the planet to cool down and accrete a significant amount of ice through accretion of comets and asteroids later on; progressively or during a "late heavy bombardment" type of event for example \citep{GLT05}. The possibility a such a scenario is supported by the fact that a debris disk has been recently imaged in the Gl\,581 system \citep{LMS12}.

\subsection{Cold trapping rate}\label{sec:trappingrate}

A constraint to the accumulation of ice from such a late water delivery is that the delivery rate must be low enough to allow trapping to take place without triggering the runaway.
To first order, our simulations presented in \sect{sec:coldtrap} can be used to quantify this cold trapping rate.

Let us consider a case where the water vapor collapses. As visible in \fig{fig:vapstart}, except for the very early times of the simulations, the water vapor content of the atmosphere tends exponentially towards its (low) equilibrium value. We can thus define an effective trapping rate $\dmvap=\mvap/(2\,\tauc)$ (in kg s$^{-1}$m$^{-2}$), where $\mvap$ is the initial water vapor column mass and $\tauc$ is the time taken by the atmosphere to decrease the water vapor amount by a factor of two. This is basically a rough estimate of the critical rate of water delivery above which runaway greenhouse would ensue.

\begin{figure}[htbp]
\begin{center}
 \resizebox{.8\hsize}{!}{\includegraphics{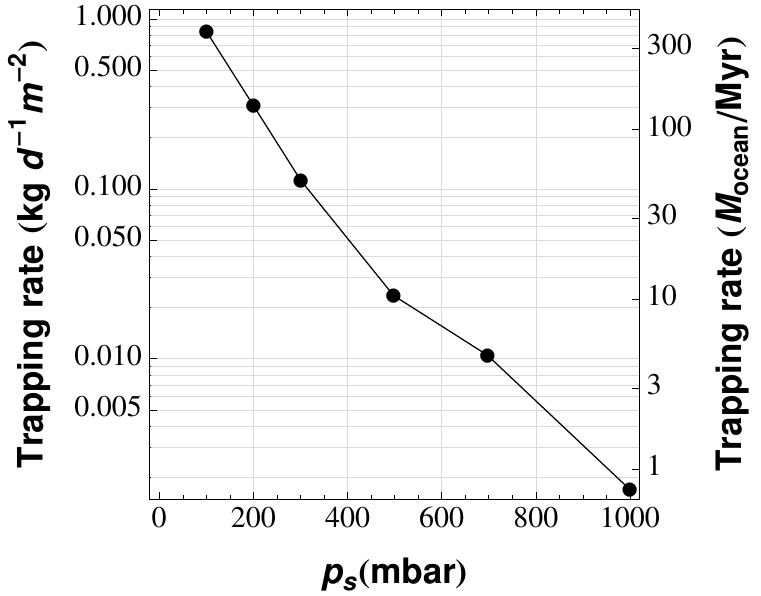}}
 \end{center}
\caption{Maximum trapping rate per unit surface ($\dmvap$) reached in our simulations as a function of the background surface pressure. The left scale shows the total trapping rate at the planetary scale in Earth oceans ($\approx1.4\times10^{21}$\,kg) per Myr.}
\label{fig:trapping}
\end{figure}

The maximum trapping rate obtained in our simulations as a function of the background atmosphere surface pressure is given in \fig{fig:trapping}. As expected, trapping is less efficient when the atmospheric mass is higher because cold traps are warmer. The exponential decrease can be understood considering that re-evaporation efficiency linearly decreases with saturation pressure that exponentially decreases with temperature following Clausius-Clapeyron law.
Then, the maximum amount of ice that can be accumulated really depends on the duration of the phase during which water is delivered to the planet, but the quite high trapping rate found in our simulations suggest that if water accretion occurs steadily, the final water inventory may not be limited by the trapping efficiency but by the rate of water delivery after the planet formation\footnote{Note that this estimation is relevant only for a synchronized planet without a thick primordial atmosphere releasing a negligible geothermal flux. All of these assumptions may be far from valid during the first few thousand to million years of the planet's life, when the planet was still releasing the energy due to its accretion and losing a possibly thick primary atmosphere of hydrogen and helium. In addition, the synchronization timescale estimated in \sect{sec:coldtrap3:2} is also of the same order of magnitude.}.


\section{On the possibility of liquid water}
\label{sec:icecap}

Our simulations show that, if the atmosphere is not too thick and the rate of water delivery high enough (while still avoiding runaway), a significant amount of water can accumulate in the cold traps. However, the low temperatures present there prevent liquid water stability. Are these land planets possibly habitable? Can liquid water flow at their surface for an extended period of time?

A possibility is that both volcanism and meteoritic impacts could be present on the night side, creating liquid water, as has been recently proposed to explain the evidences of flowing liquid water during the early Martian climate \citep{STC02,HZS07,WFM13}. Liquid water produced that way is however only episodic.

\begin{figure}[htbp]
\begin{center}
 \includegraphics[scale=.25,trim = 1cm 2.6cm 3.8cm 0cm, clip]{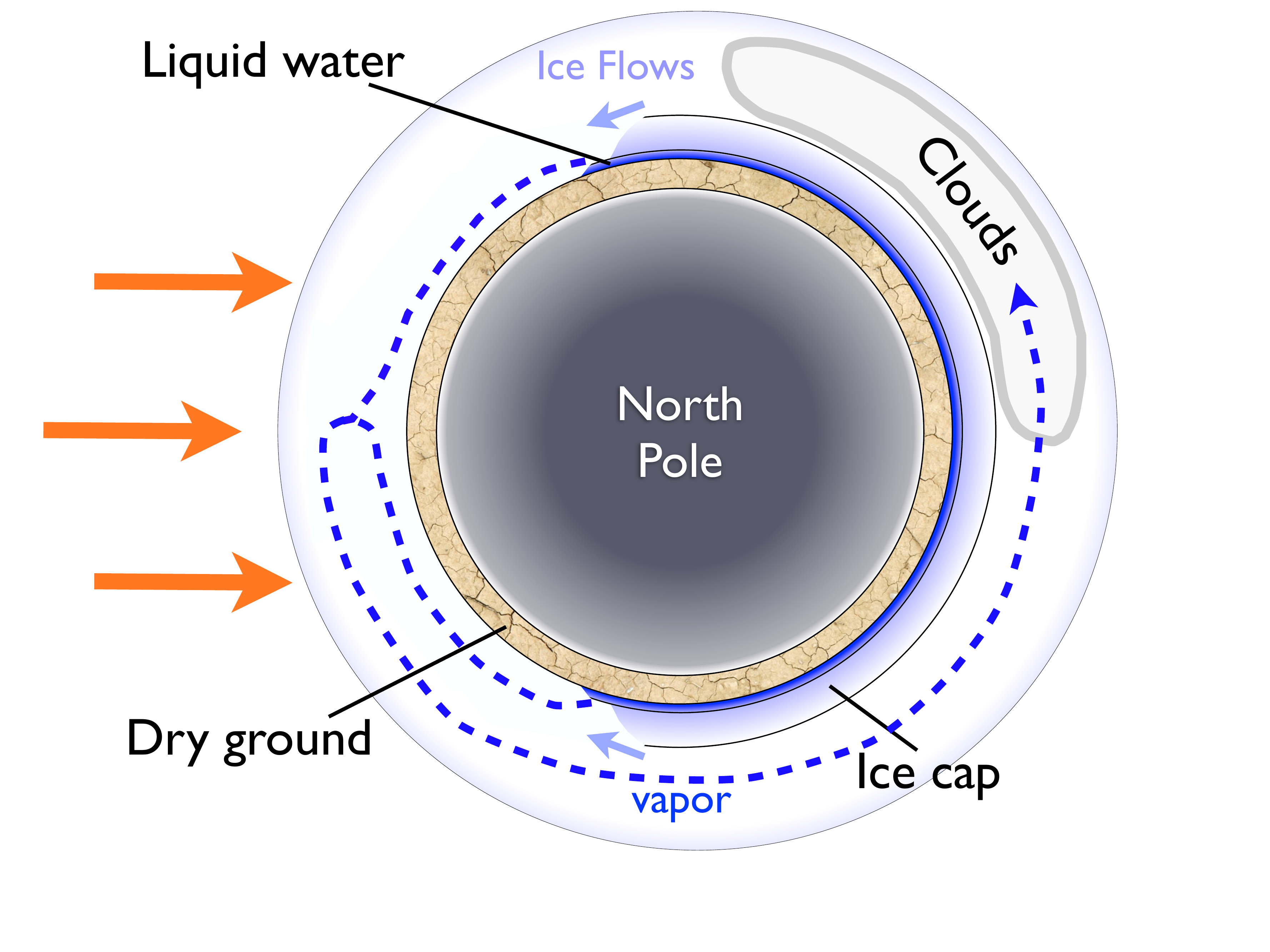}
 \end{center}
\caption{Schematic view of the the eyeball planet scenario with a dry day side and a thick ice cap on the night side.}
\label{fig:eyeball}
\end{figure}

 In this section, we briefly expose various mechanisms that could produce long-lived liquid water (as summarized in \fig{fig:eyeball}). In particular, we argue that, considering the thickness of the ice cap that could be present, gravity driven ice flows similar to those taking place on Earth could transport ice towards warm regions where it would melt. Liquid water could thus be constantly present near the edge of the ice cap. We also discuss the possibility of having subsurface liquid water produced by the high pressures reached at the bottom of an ice cap combined to a significant geothermal flux.

We also explored the quasi-snowball scenario proposed by \citet{SKL07} for \gl581 where a low equilibrium temperature is provided by a very high ice albedo ($>0.9$) and temperature slightly exceeds 0$^\circ$C at the substellar point. Notwithstanding the difficulty of having such high albedo ices around such a red star \citep{JH12}, our simulations tend to show that this scenario can be discarded. Indeed, the very strong positive feedbacks of both water vapor and ice albedo render this solution very unstable and no equilibrium state was found.

\subsection{Ice flows }

\begin{figure*}[htbp]
\begin{center}
 \subfigure{ \includegraphics[scale=.73,trim = 0cm 0.cm .65cm 0cm, clip]{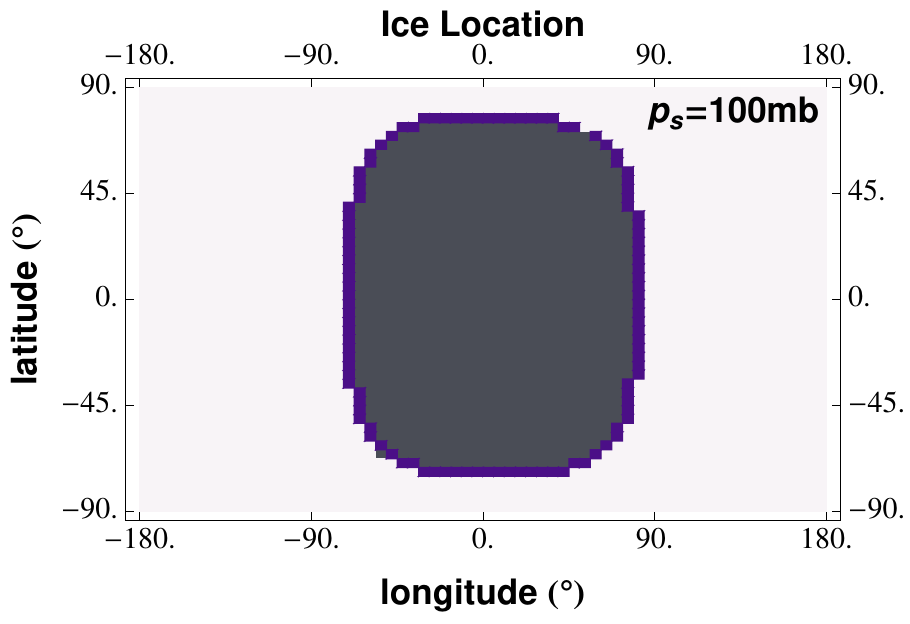} }
 \subfigure{ \includegraphics[scale=.73,trim = 1.2cm 0.cm .65cm 0cm, clip]{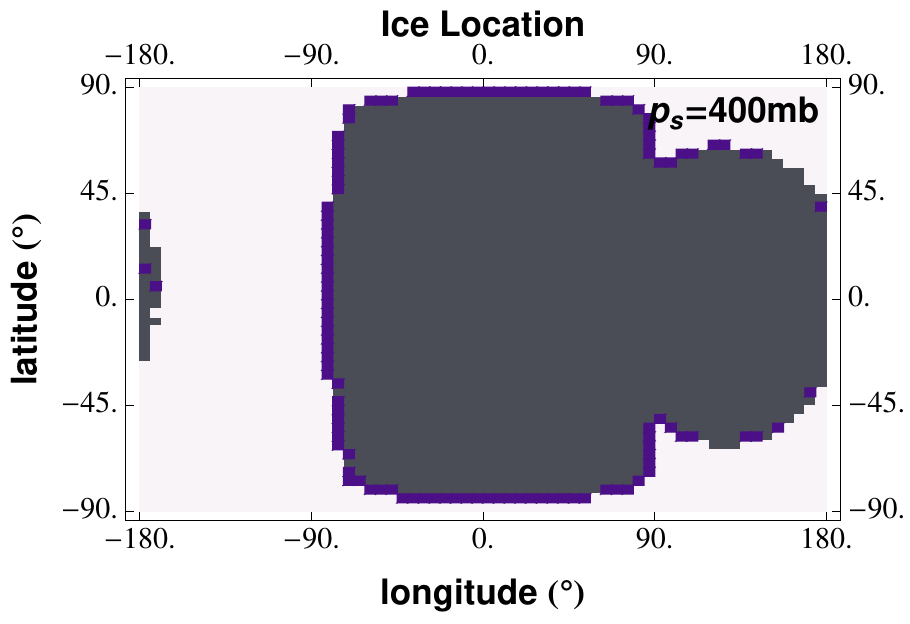} }
 \subfigure{ \includegraphics[scale=.73,trim = 1.2cm 0.cm 0cm 0cm, clip]{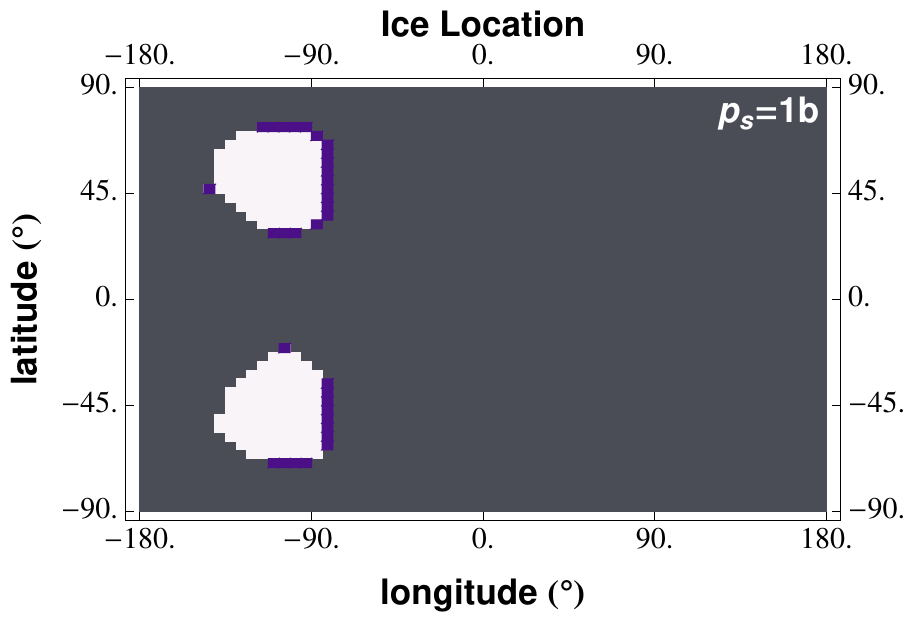} }
\end{center}
\caption{Ice distribution for different surface pressure for \gl581 ($\ps=$ 100\,mb, 400\,mb and 1\,b from left to right). Dry regions are in dark gray, ice caps are shown in white and liquid water is in blue. For background surface pressures above 300\,mb the greenhouse effect of water vapor is sufficient to melt water even on the night side.}
\label{fig:nightcap}
\end{figure*}

As described in \sect{sec:coldtrap}, if water accumulates on the surface, the atmosphere continuously acts to transport it towards the coldest regions where it is the most stable (more precipitations and less evaporation). As this accumulation seems possible mostly for low background atmospheric pressures, there always exists locations at the surface where temperature is below freezing temperature (see \figs{fig:surf_temp_map}{fig:Ts_vs_ps}) and only ice is stable.

If atmospheric transport were the only transport mechanism of water on Earth, the latter would also accumulate onto the poles and liquid water would only be present when the Sun would melt the top of the ice cap during summer. However, other processes come into play to limit the thickness of the polar ice caps. Most importantly, because ice is not completely rigid, gravity tends to drain ice from the upper regions (or regions where the ice cap is thicker) towards lower regions creating ice flows, as seen in Greenland and Antarctica among other places. These ice flows physically remove ice from the cold traps to transport it towards regions where it can potentially melt, at the edge of glaciers for example.

Assuming that a sufficient amount of water has been able to accumulate on the surface without triggering the runaway greenhouse instability, such ice flows should eventually occur. Thus, liquid water produced by melting should be present when the ice is drained to the dayside where temperatures are above 0$^\circ$C. Quantifying the amount of ice necessary and the average amount of liquid water is however much more uncertain, and maybe not relevant considering our present (lack of) knowledge concerning extrasolar planets surface properties and topography. Let it just be said that in greenland, an ice sheet of around 2\,km is sufficient to transport ice over distances of about 500-1000\,km and create ice displacements up to 2-3\,km per year near glaciers  \citep{RMS11,RM12}. As gravity (which is the main driver) and typical horizontal length scales both scale linearly with radius if density is kept constant, this order of magnitude should remain valid for more massive land planets.

As discussed in \sect{sec:ice_limits}, ice thicknesses of a few tens of kilometers could possibly be reached. It is thus possible to imagine a stationary climate state where i) a thick enough ice cap produces ice flows and liquid water at the edge of warmer regions, ii) this liquid water evaporates iii) the resulting water vapor is transported back near the coldest regions where it precipitates. A question that remains is whether or not the additional amount of water vapor that will be released in the atmosphere can be large enough to trigger the runaway greenhouse instability that would cause the melting and evaporation of the whole ice cap.

To assess the possibility of this scenario, we have run another set of simulations with a thick ice cap. Because we do not want to add any free parameter (namely the total amount of ice) and we want a worst-case scenario (the most favorable to the triggering of the runaway), we modify our base model so that the surface can act as an infinite reservoir of water whose property (liquid or solid) depends solely on the ground temperature (above or below 0$^\circ$C). In particular, we do not take into account the latent heat needed to melt water that would have a stabilizing effect on the climate by lowering the water vapor evaporation rate.

Another problem comes from the fact that freezing regions change as the amount of water vapor increases in the atmosphere. In our numerical experiment, we thus update the surface properties at each time step as follows. A dry surface grid cell becomes wet (a reservoir of water) if it has a neighbor having ice at its surface: the ice cap extends. As we want to model a case where only the edge of the ice cap melts, if a wet surface grid cell has no neighbor where water ice is present, it is dried. Then the model is initialized from the dry state from \sect{sec:dry_runs} where a seed of ice has been implanted in the cold trap, and ran until either a stationary equilibrium state is reached or the surface is completely dried by the runaway greenhouse effect. Test simulations show that the final state reached does not depend on the initial distribution of ice.

For \gl581, our simulations show that above 5\,bars, runaway is inevitable. Below 1\,bar, a stationary state is always found where ice and liquid water are both present on the surface at the same time (see \fig{fig:nightcap}). Depending on the topography near the ice cap edge, this could correspond to a situation where lakes are constantly replenished by the melting water or where rivulets moisten a more extended region. Interestingly, as shown in the middle and left panels of \fig{fig:nightcap}, ice does not necessarily cover all the nightside because of the radiative effect of water vapor which tends to warm the surface, and the inhomogeneous redistribution of energy discussed in \sect{sec:dry_runs}. In particular, melting first occurs eastward from the dayside near the equator where the surface is heated by the hot air coming from the day side through the jet, as visible in the 400\,mbar case of \fig{fig:nightcap}.

For \hd85, again, we find that runaway occurs at lower background surface pressures. The transition occurs between 400 and 500\,mb. Also, because of the more homogeneous temperature distribution of the nightside, the transition is not as progressive as shown in \fig{fig:nightcap} for \gl581. If the 400\,mb case reaches a steady state with a full ice coverage of the nightside, the 500mb case is almost dry.

These simulations demonstrate that, even in our most stringent case, in the scenario of a thick ice cap, vigorous ice flows and a significant amount of liquid water could be present without triggering the runaway greenhouse effect. Indeed, in these simulations, most of the grid cells next to the ice cap are covered by liquid water (see \fig{fig:nightcap}), and the climate remains stable. Given our resolution, this corresponds to a wet region with a width of about 1000\,km, which is probably an overestimation of the horizontal extension of the region where melting will occur. Interestingly, when there is liquid water at the surface in our model, part of the wet surface is always receiving some sunlight, because a state with liquid water only on the night side is not stable. In this case, water vapor condenses and the water freezes
. One must bear in mind that, without further information, we did not include any runoff of liquid water at the surface, assuming that the small amount of flowing water would be evaporated before crossing these 1000\,km. A more complete model of the ice sheet, ice flows and runoff would be necessary to completely settle this matter.

\subsection{Subsurface liquid water?}

Similarly to icy moons such as Europa and Enceladus \citep{SGB98,PBB99} and to subsurface lakes such as the Vostok lake on Earth \citep{OR73}, if a thick ice cap exists, the possible presence of a significant amount of subsurface liquid water at the bottom of the ice layer is also worth considering. Indeed, because of the decrease of ice melting temperature with pressure and the increase of temperature with depth caused by any geothermal flux, it is easier, in a way, to melt water below a thick ice sheet than at the surface.

To assess this possibility, let us estimate the ice thickness needed to melt water beneath it as function of the geothermal flux. Considering that vertical energy transport in the subsurface is mainly performed by thermal conduction, and that quasi-static equilibrium is reached, the temperature gradient is given by
\balign{
\dd{T}{\,z}=-\frac{\Fgeo}{\kapth},
}
where $\Fgeo$ is the (positive) geothermal flux and $\kapth$ is the thermal conductivity of water ice which ranges between 2.22\,W/m/K at 0$^{\circ}$C and 3.48\,W/m/K at -100$^{\circ}$C (citation a trouver). Thanks to hydrostatic equilibrium, and assuming a constant density and conductivity for ice (here 920\,kg/m$^3$ and 0.3\,W/m/K respectively), the temperature pressure profile is given by
\balign{\label{tpice}
T(p)=\ltsurf +\frac{\Fgeo}{\rhoi\grav\kapth}\left(p-\ps \right),
}
where $\ltsurf$ and $\ps$ are the local surface temperature and pressure and $\rhoi$ is the mass density of ice. Following \citet{WSP94}, we parametrize the melting curve of water ice by
\balign{\label{pmelt}
\pmelt(T)=\ptriple\Bigg[1 & -  0.626 \times10^6 \left(1-\left(\frac{T}{\ttriple}\right)^{-3}\right) \nonumber\\
 &+0.197135\times 10^6\left(1-\left(\frac{T}{\ttriple}\right)^{21.2}\right) \Bigg],
 }
where $\ttriple=273.16$\,K and $\ptriple=611.657$\,Pa are the temperature and pressure of the triple point of water. This parametrization is valid only between 251.165 and 273.16\,K as liquid water is not stable below this temperature at any pressure.

\begin{figure}[htbp]
\begin{center}
 \resizebox{1.\hsize}{!}{\includegraphics{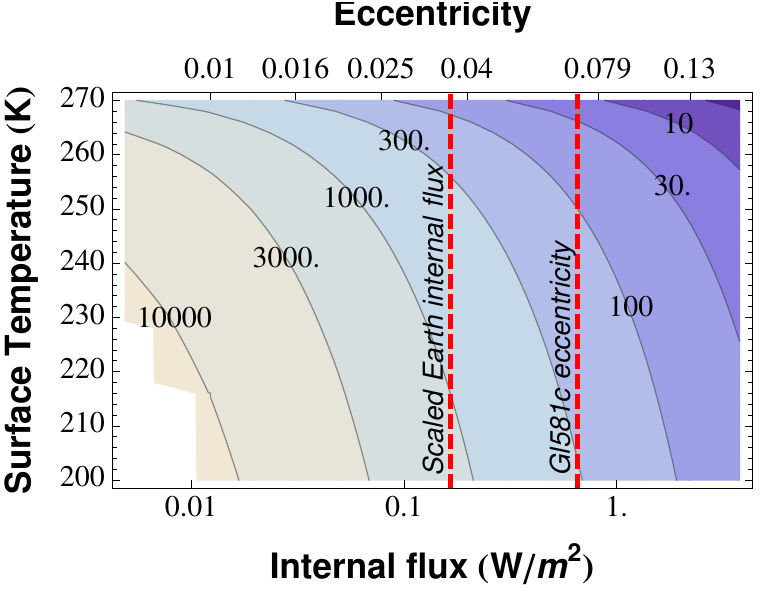}}
 \end{center}
\caption{Ice cap minimal thickness (in m) needed to reach melting temperature at the bottom of the ice sheet as a function of the temperature at its surface and internal heat flux. The white area represents the region where conditions are too cold for liquid water to exist, even at high pressure. This internal flux could be caused by release of primordial heat content, radioactive decay and tidal heating. The top scale shows the eccentricity needed to provide the given flux for \gl581 assuming Earth tidal response (see text).}
\label{fig:icecap_min_thickness}
\end{figure}

Then, for a given geothermal flux and surface temperature, subsurface liquid water can exist if the curves parametrized by \eqs{tpice}{pmelt} intersect each other.
By solving this implicit equation, we can find the pressure at which melting occurs and define the minimal ice cap thickness needed to reach this pressure, $l_\mathrm{ice}=(p-\ps)/(\rhoi\grav)$ (for the atmosphere considered, $\ps$ can safely be neglected in the above equations). Numerical estimates of this thickness as a function of surface temperature and geothermal flux are shown in \fig{fig:icecap_min_thickness} for a typical gravity of 18\,m/s$^2$.

One can see that even if no geothermal flux is present, subsurface liquid water could exist for warm enough regions. The thickness needed would however be rather large ($\sim$3-10\,km) which seems unlikely. Depending on the amount of internal flux, however, the minimal thickness can be much lower. On Earth, radioactive decay and release of primordial heat generate a globally averaged geothermal flux of 92\,mW/m$^2$ \citep{DD10}. If we assume that the rate of volumetric heating is constant, this mean flux should scale linearly with the planet's radius, and our more massive planet should produce between 1.6 and 1.8 times this flux. With such a flux, the thickness needed decreases below 1.5\,km in the coldest regions but could be as low as 300\,m.

\begin{figure}[htbp]
\begin{center}
  \resizebox{1.\hsize}{!}{\includegraphics{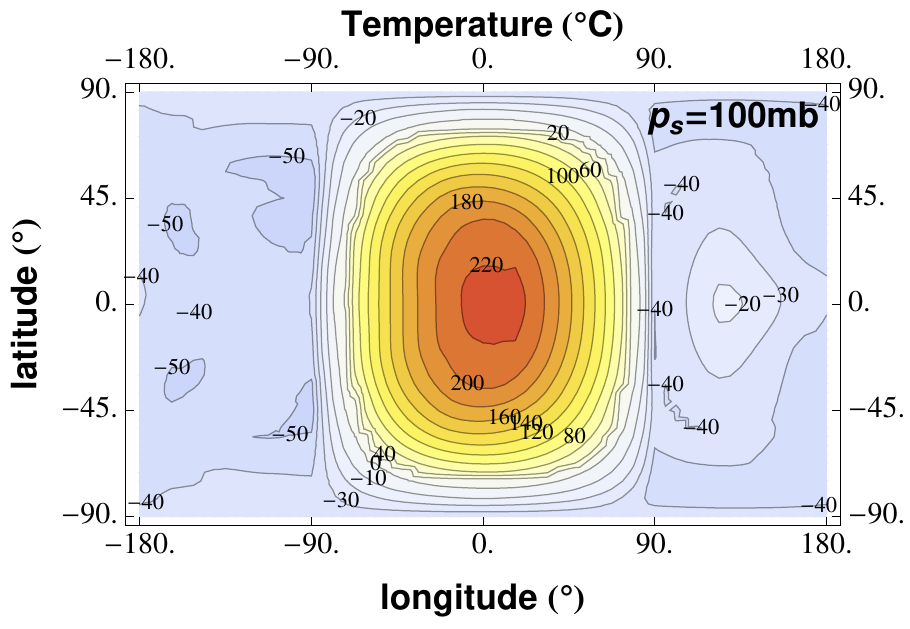}}
\end{center}
\caption{Surface temperature map (in $^{\circ}$C) in the 100\,mb case in presence of a thick ice cap and liquid water near the terminator (see text). Dayside temperatures are up to 50\,K higher than in the dry case (upper left panel of \fig{fig:surf_temp_map}).
}
\label{fig:nightcap_ts}
\end{figure}

Finally, another important source of internal flux for close in planets is the friction created by the bodily tides raised by the host star. An estimate of this tidal dissipation in the synchronous case is given by Eq.\,(14) of \citet{LCB10} (see also \citealt{hut81}). This heating is a function of the orbital eccentricity of the planet and of the efficiency of the tidal dissipation into the planet's interior (the imaginary part of the Love number, $\kp\,\dtp$; see \sect{sec:coldtrap3:2}). For an Earth-like dissipation and eccentricities shown in \tab{tab:params}, the flux can be estimated to be $\sim$ 600\,mW/m$^2$ for \gl581 and would contribute in a large part to the interior energy budget\footnote{Note that this estimation differs from the one of \citet{SKL07} because they used an eccentricity of 0.16 as initially found by \citet{UBD07}.}. As shown in \fig{fig:icecap_min_thickness}, the thickness needed to produce liquid water would be much lower ($<300\,$m). Because the measured eccentricity is highly uncertain, we computed the eccentricity needed to produce the internal fluxes shown in \fig{fig:icecap_min_thickness} (top scale of the figure). However, one must bear in mind that the dissipation constant is also highly uncertain, and the tidal flux computed here could easily change by an order of magnitude. For \hd85 the tidal flux is much lower (0.1\,mW/m$^2$) because the planet is much further away from its parent K star.

Extended reservoirs of subsurface water can thus not be ruled out. The ice thickness needed, if realistic internal fluxes are present, is much lower than the one that could potentially be accumulated as discussed in \sect{sec:ice_limits}.

\section{Observable signature}\label{sec:obs}


\begin{figure}
\floatbox[{\capbeside\thisfloatsetup{capbesideposition={right,bottom},capbesidewidth=4cm}}]{figure}[\FBwidth]
{\caption{Bond albedo as a function of water vapor column amount for \gl581 runs with $\ps=200$\,mb (gray dashed curve) and $\ps=700$\,mb (black solid curve). The surface albedo of 0.3 is shown as the long dashed horizontal line.}\label{fig:alb_vs_h2o_vap}}
{\includegraphics[scale=.9]{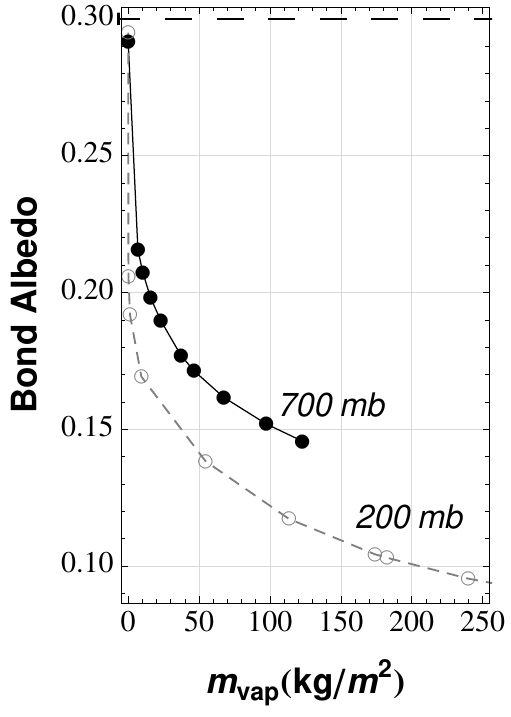}}
\end{figure}

\begin{figure*}[tbp]
\begin{center}
\subfigure{ \includegraphics[scale=1.0,trim = 0cm 0cm 0cm 0cm, clip]{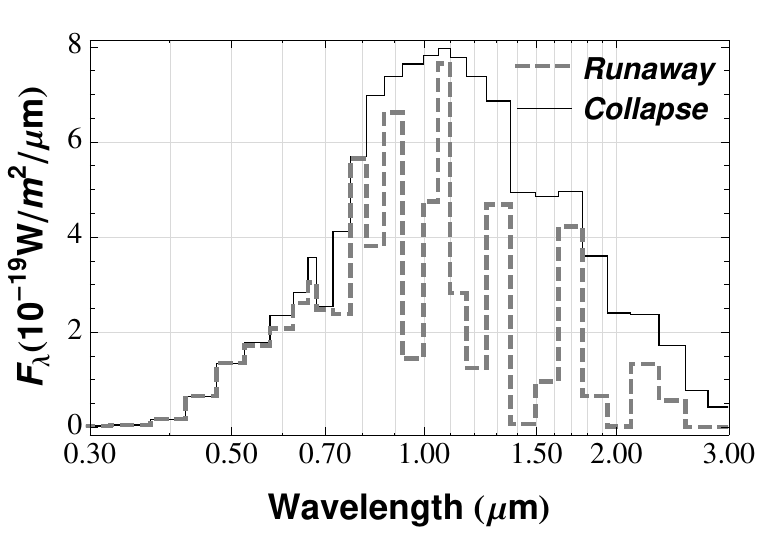}}
\subfigure{ \includegraphics[scale=1.,trim = 0cm 0cm 0cm 0cm, clip]{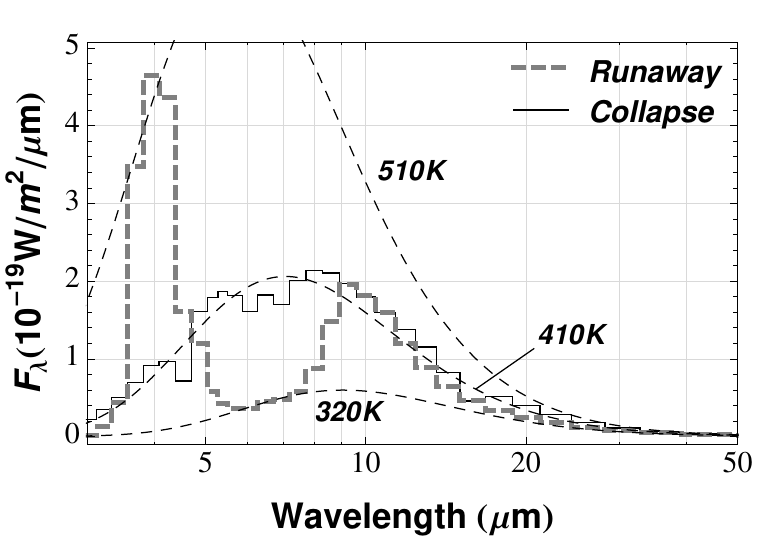}}
\end{center}
\caption{Synthetic spectrum in reflection (left) and thermal emission (right) during secondary transit of a \gl581 like planet with a 200\,mb atmosphere at 10\,parsecs.
The black solid and the gray dashed lines respectively stand for the dry collapsed and runaway case presented in \fig{fig:TS_runaway_collapse}. Thin dashed lines in the emission part correspond to blackbody spectra at the temperatures indicated. In the runaway case water is abundant in the atmosphere and H2O absorption bands are noticeable throughout the spectrum (between 5 and 9\,$\mu$m in the emission spectrum and above 0.7\,$\mu$m in reflection). The shift of the thermal emission toward shorter wavelengths due to the much hotter surface is also striking.}
\label{fig:spectrum}
\end{figure*}

Can observation allow us to distinguish between the different climate regimes presented here? To start answering this question, let us consider various possible observables.

If a given target's flux is too small to use spectroscopy, broad band photometry (as has been done for \textit{Kepler}-10\,b \citep{BBB11} and \textit{CoRoT}-7\,b \citep{LRS09}) can already provide constraints. Indeed, as shown in \fig{fig:alb_vs_h2o_vap}, the presence of water vapor in the atmosphere of a planet orbiting a M dwarf strongly decreases its Bond albedo because many water spectral bands peak at the same wavelengths than the stellar spectrum. However, while this method can hint at the presence of a strong absorber in the atmosphere of the planet observed - hinting at the presence of an atmosphere itself - a strong degeneracy exist between the various possible gases and spectroscopy is needed. This signature will of course decrease for more massive, bluer stars.

To go further, we have computed synthetic spectra of various models depicted above.
We have used the tool developed by \citet{SWF11} that can model the spectral phase curve of a planet in an arbitrary geometrical configuration directly from the output of the GCM code. Here we focus only on the spectrum that could be observed during the secondary transit of a transiting analogue of \gl581 at 10\,parsecs.
An example of the spectra computed for the 200\,mb runaway and collapse cases presented in \figs{fig:vapstart}{fig:TS_runaway_collapse} are shown in \fig{fig:spectrum}.

Although the intensity of the bolometric signal is weak, on can see that spectral differences between a runaway and a collapse case are striking. In reflection first, on can easily spot the absorption bands of water vapor in the runaway case which are not present in the dry, collapse case. Indeed, when water vapor becomes a significant constituent of the atmosphere, the latter becomes thick in the \h2o molecular bands even in the near infrared. A large fraction of the incoming stellar radiation is thus absorbed in those bands before being reflected by the surface or Rayleigh scattering.
Secondly, in the emission spectrum, one can see that the planet in runaway can only radiate trough the \h2o window regions around 4 and 10\,$\mu$m. The surface that is heated by the strong resulting greenhouse effect thus radiates at shorter wavelengths.

Further observation may thus, if a sufficient accuracy is met (see \citet{SWF11} for a more detailed discussion), provide strong constraints on the climate regime reached on such strongly irradiated land planets as the presence of strong \h2o spectral features should be correlated with shape of the thermal emission.

\section{What's next?}\label{sec:conc}

As for \gl581 or \hd85, the habitability of planets receiving more flux than the classical runaway greenhouse has often been discarded on the basis of 1D models that predict that an ocean cannot be stable under these conditions.

Here, we have demonstrated that because of the inhomogeneity of the impinging flux and of the natural ability of the atmosphere to transport condensable species toward colder regions where it can be captured, presence of surface water cannot be ruled out on the sole basis of the mean absorbed stellar flux. In fact, because of the strong positive feedbacks of water vapor, two stable equilibrium climate regimes exist for the atmosphere, a runaway greenhouse state and a state where all the water is trapped on the surface in cold regions. A key finding of this study is that the state in which the atmosphere settles depends not only on the incoming flux, but on many factors: atmospheric mass, water content and initial distribution, spin state, and circulation regime.

If it is not easy to set clear limits on the domain of existence of this bistable regime, however, a few trends emerge. First, the mechanism described above strongly relies on the existence of an efficient permanent (or slowly evolving) cold trap. The trapping in a spin orbit resonance or a very low obliquity thus seem to strongly favor the onset of the bistability. Second, except for changes in the circulation pattern, the domain of existence of the collapse state narrows down in the pressure/initial water vapor amount parameter space when the incoming flux increases. These two argument suggest that the domain of the bistability should become smaller when the star becomes more massive because regions where tidal interactions causing resonances to occur and damping obliquity move at higher fluxes. Again, it is difficult to set a clear limit in terms of a spectral type above which this bistability cannot occur. Let us just say that our simulations support the existence of a bistability around M and K dwarfs, and that, if not for its thick atmosphere, Venus climate would be very close to the one of \hd85, extending this statement to G dwarfs. The presence of ice on the poles of Mercury \citep{LFG13} also bring observational support to this scenario, at least in the limit of a vanishing atmosphere. The validity of our results thus seems to extend to a broad range of situations.

Interestingly, because all the parameters affecting the occurrence of a bistable state change during the life of the planet, the present climate of a given object may depend on the evolutionary path that it followed. In particular, the history of the water delivery and of the atmospheric escape may play a crucial role in determining if these objects are dry rocks surrounded by a hot steam atmosphere or harbor vast ice deposits on the night side and near the poles.

Another finding is that the atmospheric circulation present on synchronous extrasolar planets does not necessary settles in a state of super-rotation. As the super-rotation seems to be powered by planetary scale Rossby and Kelvin waves, we find that it weakens when the wavelength of these waves, i.e. the equatorial Rossby deformation radius, significantly exceeds the planetary radius. For small and/or very slowly rotating planets (synchronously rotating on a wide orbit for example) the circulation seems to be dominated by a symmetrical stellar to anti stellar circulation at high altitudes.

To go further ahead, observational data are needed.
Fortunately, exoplanet-characterizing observatories such as EChO \citep{TBH12} and JWST should soon come in line. And even if the characterization of Earth analogs around M dwarfs will prove very challenging, we will learn a lot from the observation of planets beyond the runaway greenhouse limit such as transiting versions of the ones described here and hotter. Combining spectral observations of the primary and secondary transits (and possibly phase curves) of a number of strongly irradiated terrestrial exoplanets should indeed help us to unravel the variety of atmosphere masses, compositions and histories; so far the least constrained parameters used in exoplanetary atmospheres modeling.

\begin{acknowledgements}
JL thanks V. Makarov and M. Efroimsky for providing calculations on the spin state of \gl581 and O. Grasset for an inspiring discussion concerning icy moons and subsurface oceans. Many thanks to A. Spiga for his thorough reading of the manuscript. We thank our referee, A. Showman, for his insightful comments.
JL received funding from the DIM ACAV. F. S. acknowledges support from the European Research Council (ERC Grant 209622: E$_3$ARTHs).
\end{acknowledgements}

\bibliography{biblio}
\bibliographystyle{aa}



\end{document}